\documentclass[11pt]{article}
\usepackage{amssymb,amsmath,amsfonts}
\usepackage{graphicx}
\usepackage{graphics}
\usepackage{eepic,epsfig}

\makeatletter
\@addtoreset{equation}{section}

\makeatother

\begin{document}

\title{Fermionic condensate and the mean energy-momentum tensor \\
in the Fulling-Rindler vacuum}
\author{S. Bellucci$^{1}$\thanks{%
E-mail: Stefano.Bellucci@lnf.infn.it },\thinspace\ V. Kh. Kotanjyan$^{2,3}$%
\thanks{%
E-mail: vatokoto@gmail.com},\thinspace\ A. A. Saharian$^{2}$\thanks{%
E-mail: saharian@ysu.am} \\
\\
\textit{$^1$ INFN, Laboratori Nazionali di Frascati,}\\
\textit{Via Enrico Fermi 54, 00044 Frascati (Roma), Italy} \vspace{0.3cm}\\
\textit{$^2$Institute of Physics, Yerevan State University,}\\
\textit{1 Alex Manoogian Street, 0025 Yerevan, Armenia} \vspace{0.3cm}\\
\textit{$^3$Institute of Applied Problems of Physics NAS RA,}\\
\textit{25 Hrachya Nersissyan Street, 0014 Yerevan, Armenia}}
\maketitle

\begin{abstract}
We investigate the properties of the fermionic Fulling-Rindler vacuum for a
massive Dirac field in a general number of spatial dimensions. As important
local characteristics, the fermionic condensate and the expectation value of
the energy-momentum tensor are evaluated. The renormalization is reduced to
the subtraction of the corresponding expectation values for the Minkowski
vacuum. It is shown that the fermion condensate vanishes for a massless
field and is negative for nonzero mass. Unlike the case of scalar fields,
the fermionic vacuum stresses are isotropic for general case of massive
fields. The energy density and the pressures are negative. For a massless
field the corresponding spectral distributions exhibit thermal properties
with the standard Unruh temperature. However, the density-of-states factor
is not Planckian for general number of spatial dimensions. Another
interesting feature is that the thermal distribution is of the Bose-Einstein
type in even number of spatial dimensions. This feature has been observed
previously in the response of a particle detector uniformly accelerating
through the Minkowski vacuum. In an even number of space dimensions the
fermion condensate and the mean energy-momentum tensor coincide for the
fields realizing two inequivalent irreducible representations of the
Clifford algebra. In the massless case, we consider also the vacuum
energy-momentum tensor for Dirac fields in the conformal vacuum of the Milne
universe, in static open universe and in the hyperbolic vacuum of de Sitter
spacetime.
\end{abstract}

\bigskip

\section{Introduction}

The observer dependence of the vacuum and particle notions is among the
important lessons from quantum field theory on curved spacetimes. The
crucial point in the quantization procedure is the choice of the complete
set of mode functions for a given field. The expansion of the field operator
over those modes determines the annihilation and creation operators. The
construction of the Fock space of states starts from the definition of the
vacuum state that is nullified by the action of the annihilation operator.
The particle is a state of quantum field obtained acting by the creation
operator on the vacuum state. It carries a set of quantum numbers that
determines the mode functions of the classical field equations. From this
construction it follows that the vacuum and particle states, in general,
depend on the choice of the mode functions. The annihilation and creation
operators for different sets of mode functions are related by the Bogoliubov
transformations. If those transformations mix the annihilation and creation
operators, the Bogoliubov $\beta $-coefficient is different from zero and
the two vacuum states based on two sets of modes are not equivalent: the
vacuum state corresponding to one set of modes contains particles of the
other set of modes.

The natural mode functions used in the expansion of the field operator may
differ for different observers, giving rise to different vacuum states. This
may take place already in flat spacetime. The classical example of two
inequivalent vacuum states in the Minkowski spacetime are the Minkowski and
Fulling-Rindler vacua. They are vacuum states for inertial and uniformly
accelerating observers, respectively. The quantization of fields in Rindler
coordinates, that are the natural coordinates for uniformly accelerating
observers, has been widely discussed in the literature (see \cite%
{Birr82,Cris08,Lang05} and references therein). The interest is motivated by
several reasons. First of all, it comes from principal questions of
quantization of fields in geometries having horizons. The latter can be
either observer dependent (like Rindler or de Sitter (dS) horizons) or
determined by the matter distribution (examples are the black hole
horizons). The Rindler geometry is simple enough to allow to find exact
solutions in different problems of quantum field theory. This may shed light
on the respective problems in more complicated geometries where the exact
solutions are not available or they are complicated. Next, the Rindler
metric approximates the black hole geometry in the near horizon limit and
the roots of a number of quantum field theoretical phenomena around black
holes can be found in the Rindler physics. An example is the relation
between the Unruh effect and Hawking radiation. The Unruh effect \cite%
{Full73,Davi75,Unru76} states a kind of equivalence between thermal
fluctuations of a quantum field observed by an inertial observer and
fluctuations in the inertial (Minkowski) vacuum of the same field recorded
by a uniformly accelerating observer. The temperature of the thermal bath,
the Unruh temperature, is proportional to the acceleration of the Rindler
observer. The thermal nature of the distribution of Rindler particles is
closely related to the presence of the horizon for uniformly accelerating
observers. The events outside the horizon are not accessible for those
observers and they are traced out, resulting in the information loss and
thermal state. Being a background with horizons, the Rindler geometry is an
interesting arena to investigate the phenomena of quantum entanglement (see,
e.g., Refs. \cite{Fuen05,Alsi06,Ueda21,Yan22}). From the point of view of a
uniformly accelerating observer the Minkowski vacuum appears as an entangled
state between the states in the right and left wedges of the Rindler
decomposition in the Minkowski spacetime. The considerations of specific
examples have shown that the Unruh effect can either reduce or enhance
entanglement. Though the experimental observation of the Unruh effect
requires huge accelerations, different schemes for simulating the effect in
laboratory have been discussed in the literature (see, for example, \cite%
{Cris08, Mart11,Rodr17} and references therein).

The investigations in the Rindler physics were carried out in two main
directions. The first one considers the properties of the Minkowski vacuum
and of the related particle states seen by Rindler observers. In particular,
a large number of papers are devoted to the study of the response of various
types of particle detectors in an accelerated motion through the inertial
vacuum (for different types of particle detectors interacting with quantum
fields see, for example, \cite{Lang05,Humm16,Sach17}). The main subject in
the second class of investigations are the properties of the Fulling-Rindler
vacuum and of the Rindler particle states from the viewpoints of both
Rindler and inertial observers. For free quantum fields, among the important
local characteristics of the vacuum state are the expectation values of
bilinear products of the field operator, like the field squared and the
energy-momentum tensor. The present paper deals with those characteristics
of the fermionic Fulling-Rindler vacuum state for a massive Dirac field in
general number of spatial dimensions. Various aspects of quantum fermionic
fields in Rindler spacetime have been considered in the literature. The
references \cite{Alsi06}-\cite{Falc23} include an incomplete list of some of
them. Our consideration of general number of spatial dimension is motivated
by possible applications in high energy models of fundamental physics such
as string theories, supergravities, Kaluza-Klein type theories and
braneworld models. An interesting application of fermionic models in
two-dimensional space comes from condensed matter physics. In the long
wavelength approximation the excitations of the electronic subsystem in
so-called Dirac materials are well described by the Dirac model in two
dimensions where the speed of light is replaced by the Fermi velocity (see
reviews \cite{Gusy07,Cast09}). The latter is smaller than the speed of light
by orders of magnitude and the Dirac materials provide a unique possibility
for investigations of relativistic effects at smaller velocities. An example
of such a condensed matter system is graphene. The graphene based
structures, like carbon nanotubes and nanoloops, also provide an opportunity
to study the effects of nontrivial spatial topology on the properties of the
fermionic vacuum in quantum field theory \cite{Bell09}-\cite{Bell13}.
Effects of compactification of spatial dimensions in the Fulling-Rindler
vacuum for a scalar field have been recently discussed in \cite{Kota22}.

The remainder of the paper is structured as follows. In the next section we
present the positive and negative energy Rindler modes for a massive Dirac
field in general number of spatial dimensions. By using those modes, in
Section \ref{sec:FC} the renormalized fermion condensate is evaluated. The
renormalization is based on the subtraction of the corresponding vacuum
expectation value (VEV) for the Minkowski vacuum state. The corresponding
considerations for the vacuum expectation value of the energy-momentum
tensor are presented in Section \ref{sec:EMT}. We show that, as expected,
the mean energy-momentum tensor obeys the covariant conservation equation
and the trace relation. Alternative representations for the fermion
condensate and vacuum energy-momentum tensor, well adapted for numerical
evaluations, are derived in Section \ref{sec:AlterRep}. In particular, they
show that the vacuum stresses are isotropic for the Dirac field. Numerical
results are presented as well. Section \ref{sec:Conc} concludes the main
results of the paper. In Appendix \ref{sec:App1}, integral representations
are provided for products of the modified Bessel functions with imaginary
order. Those representations have been used to simplify the expressions for
the VEVs. In Appendix \ref{sec:AppMink} we give an integral representation
for the Hadamard function of the Dirac field in the Minkowski vacuum that is
used in the subtraction procedure for renormalized VEVs. An integral
representation for the difference of the traces for the Hadamard functions
in the Fulling-Rindler and Minkowski vacua is provided in Appendix \ref%
{sec:App3}.

\section{Fermionic modes in Rindler spacetime}

\label{sec:Modes}

The complete set of fermionic modes are required in the canonical
quantization procedure. In the literature they are mainly considered in
spatial dimensions $D=1$ and $D=3$ (for a recent discussion of solutions to
the Dirac equation in 4-dimensional Rindler spacetime with analytic
continuations to all quadrants of Minkowski spacetime see \cite%
{Ueda21,Falc23} and references therein). Here, we generalize the approach of
Ref. \cite{Cand78} for a massive field and for general number of spatial
dimension $D$ and present the complete set of Dirac modes in the form that
does not depend on the special representation of the Dirac matrices. They
are well adapted for investigations of the VEVs of physical observables.

The dynamics of a fermionic field $\psi (x)$, $x=(x^{0},x^{1},\ldots ,x^{D})$%
, in a $(D+1)$-dimensional spacetime with the metric tensor $g_{\mu \nu }(x)$
is described by the Dirac equation
\begin{equation}
\left( i\gamma ^{\mu }\nabla _{\mu }-m\right) \psi =0,  \label{eom}
\end{equation}%
where $\gamma ^{\mu }=e_{(b)}^{\mu }\gamma ^{(b)}$ are the curved spacetime
gamma matrices, $\nabla _{\mu }=\partial _{\mu }+\Gamma _{\mu }$ is the
covariant derivative for Dirac fields and $\Gamma _{\mu }$ is the spin
connection. The expression for the latter in terms of the flat spacetime
matrices $\gamma ^{(a)}$, $a=0,1,\ldots ,D$, and $(D+1)$-bein fields $%
e_{(a)}^{\mu }$ reads
\begin{equation}
\Gamma _{\mu }=\frac{1}{4}\gamma ^{(a)}\gamma ^{(b)}e_{(a)}^{\nu }e_{(b)\nu
;\mu },  \label{Gammu}
\end{equation}%
with the semicolon meaning the standard covariant derivative of vector
fields. We will consider a fermionic field realizing the irreducible
representation of the Clifford algebra $\left\{ \gamma ^{\mu },\gamma ^{\nu
}\right\} =2g^{\mu \nu }$ with $N\times N$ Dirac matrices. Here, $%
N=2^{[(D+1)/2]}$ and $[z]$ is the integer part of $z$ (for the Dirac
matrices in an arbitrary number of the spacetime dimension see, for example,
\cite{Shim85,Park09}). Up to a similarity transformation, the irreducible
representation is unique for odd $D$. In the case of even $D$ there are two
inequivalent irreducible representations.

\begin{figure}[tbph]
\begin{center}
\epsfig{figure=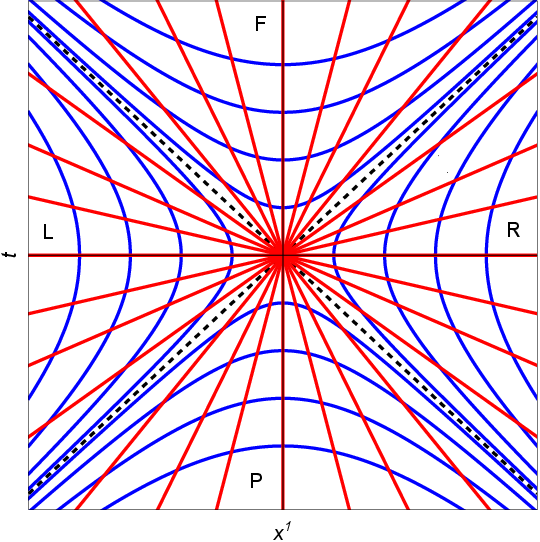,width=7.5cm,height=7.5cm}
\end{center}
\caption{The subspace $(t,x^{1})$ of Minkowski spacetime covered by Rindler
coordinates $(\protect\tau ,\protect\rho )$. The four wedges, the R, L, F,
and P regions, are separated by the Rindler horizons $t=\pm x^{1}$ (dashed
lines).}
\label{Coorfig}
\end{figure}

The background geometry under consideration is described by the Rindler line
element%
\begin{equation}
ds_{\mathrm{R}}^{2}=g_{\mu \nu }dx^{\mu }dx^{\nu }=\rho ^{2}d\tau ^{2}-d\rho
^{2}-d\mathbf{x}^{2},  \label{ds2}
\end{equation}%
where $\mathbf{x}=\left( x^{2},x^{3},\ldots ,x^{D}\right) $, $-\infty <\tau
<+\infty $, and $0\leq \rho <\infty $. By the coordinate transformation
\begin{equation}
t=\rho \sinh \tau ,\;x^{1}=\rho \cosh \tau ,  \label{trans}
\end{equation}%
the line element takes the Minkowskian form $ds_{\mathrm{M}}^{2}=dt^{2}-d%
\mathbf{z}^{2}$, $\mathbf{z}=(x^{1},\mathbf{x})$, and the geometry is flat.
The worldline with fixed spatial coordinates $(\rho ,\mathbf{x})$
corresponds to a uniformly accelerating observer moving along the line
parallel to the $x^{1}$ axis with proper acceleration $1/\rho $. The proper
time of that observer is measured by $\tau _{p}=\rho \tau $. We note that
the Rindler coordinates divide the subspace $(t,x^{1})$ of the Minkowski
spacetime into four regions (wedges) depicted in Figure \ref{Coorfig}. The
transformation (\ref{trans}) corresponds to the right wedge (R region) with $%
x^{1}>|t|$. The transformation presenting the left wedge (L region), with $%
x^{1}<-|t|$, is obtained from (\ref{trans}) adding the minus sign in the
right-hand side for the expression of $x^{1}$. The coordinate
transformations for the remaining regions, future (F) and past (P) regions
with $t>|x^{1}|$ and $t<-|x^{1}|$, respectively, are given by $t=\pm \rho
\cosh \tau $ and$\;x^{1}=\rho \sinh \tau $. Here, the upper and lower signs
correspond to the F and P regions. In the discussion below we will consider
the VEVs in the R region. The same expressions for the VEVs are obtained in
the L region.

We want to find a complete set of solutions to equation (\ref{eom}) for the
geometry (\ref{ds2}). The corresponding metric tensor is given by%
\begin{equation}
g_{\mu \nu }=\mathrm{diag}(\rho ^{2},-1,-1,\ldots ,-1).  \label{gmunu}
\end{equation}%
The $(D+1)$-bein fields can be chosen as $e_{(0)}^{\mu }=\delta _{0}^{\mu
}/\rho $ and $e_{(b)}^{\mu }=\delta _{b}^{\mu }$ for $b=1,2,\ldots $. The
related spin connection is expressed as
\begin{equation}
\Gamma _{\mu }=\frac{1}{2}\gamma ^{(0)}\gamma ^{(1)}\delta _{\mu }^{0}.
\label{Gamu}
\end{equation}%
We present the solution of the equation (\ref{eom}) in the form%
\begin{equation}
\psi =\left( i\gamma ^{\nu }\nabla _{\nu }+m\right) \varphi (x),
\label{Anzats}
\end{equation}%
with a new field $\varphi (x)$. For the latter the following equation is
obtained%
\begin{equation}
\left( g^{\mu \nu }\nabla _{\mu }\nabla _{\nu }+m^{2}\right) \varphi (x)=0,
\label{Equ}
\end{equation}%
with the same operator $\nabla _{\mu }$ as in (\ref{eom}).

By taking into account (\ref{Gamu}), for the geometry at hand the field
equation (\ref{Equ}) is reduced to%
\begin{equation}
\left( \partial _{\tau }^{2}+\gamma ^{(0)}\gamma ^{(1)}\partial _{\tau
}-\rho ^{2}\partial _{\rho }^{2}-\rho \partial _{\rho }-\rho
^{2}\sum_{l=2}^{D}\partial _{l}^{2}+\frac{1}{4}+m^{2}\rho ^{2}\right)
\varphi (x)=0.  \label{Equ2}
\end{equation}%
The geometry possesses a Killing vector $\partial _{\tau }$ and the solution
of Eq. (\ref{Equ2}) corresponding to the positive energy fermionic modes
with respect to this vector can be presented in the form
\begin{equation}
\varphi (x)=u(\rho ,\omega ,\mathbf{k})e^{-i\omega \tau +i\mathbf{k}\cdot
\mathbf{x}},  \label{usol}
\end{equation}%
where $\mathbf{k}=\left( k^{2},k^{3},\ldots ,k^{D}\right) $, $-\infty
<k^{l}<+\infty $, $l=2,\ldots ,D$, $\mathbf{k}\cdot \mathbf{x}$ $%
=\sum_{l=2}^{D}k^{l}x^{l}$, and $0\leq \omega <\infty $. Plugging this in (%
\ref{Equ2}), we get the equation for the function $u(\rho ,\omega ,\mathbf{k}%
)$:%
\begin{equation}
\left[ \partial _{\rho }^{2}+\frac{1}{\rho }\partial _{\rho }-\lambda ^{2}+%
\frac{1}{\rho ^{2}}\left( \omega ^{2}+i\gamma ^{(0)}\gamma ^{(1)}\omega -%
\frac{1}{4}\right) \right] u(\rho ,\omega ,\mathbf{k})=0,  \label{ueq2}
\end{equation}%
with $k=|\mathbf{k}|$ and
\begin{equation}
\lambda =\sqrt{k^{2}+m^{2}}.  \label{lam}
\end{equation}%
By taking into account that%
\begin{equation}
\omega ^{2}+i\omega \gamma ^{(0)}\gamma ^{(1)}-\frac{1}{4}=-\left( i\omega -%
\frac{1}{2}\gamma ^{(0)}\gamma ^{(1)}\right) ^{2},  \label{rel1}
\end{equation}%
the solution of (\ref{ueq2}), finite in the limit $\rho \rightarrow \infty $%
, is expressed as%
\begin{equation}
u(\rho ,\omega ,\mathbf{k})=K_{i\omega -\frac{1}{2}\gamma ^{(0)}\gamma
^{(1)}}\left( \lambda \rho \right) \chi (\mathbf{k}),  \label{u3}
\end{equation}%
where $K_{\nu }(z)$ is the modified Bessel function of the second kind
(Macdonald function) and $\chi (\mathbf{k})$ is a constant spinor. Note that
for a function $f(x)$ with the argument $x=\gamma ^{(0)}\gamma ^{(1)}$ we
have (see also \cite{Cand78})%
\begin{equation}
f(\gamma ^{(0)}\gamma ^{(1)})=\frac{1}{2}\sum_{\varkappa =\pm 1}\left(
1+\varkappa \gamma ^{(0)}\gamma ^{(1)}\right) f(\varkappa ).  \label{rel2}
\end{equation}

Having the function $\varphi (x)$, for the solution corresponding to the
Dirac spinor $\psi (x)$ one finds%
\begin{equation}
\psi =-2i\lambda e^{-i\omega \tau +i\mathbf{k}\cdot \mathbf{x}}K_{i\omega -%
\frac{1}{2}\gamma ^{(0)}\gamma ^{(1)}}\left( \lambda \rho \right) P(\mathbf{k%
})\gamma ^{(1)}\chi (\mathbf{k}),  \label{psi2}
\end{equation}%
where we have introduced the notation%
\begin{equation}
P(\mathbf{k})=\frac{1}{2}\left( 1-i\gamma ^{(1)}\frac{\boldsymbol{\gamma \,}%
\mathbf{k}+m}{\lambda }\right) ,  \label{Pk}
\end{equation}%
with $\boldsymbol{\gamma \,}\mathbf{k}=\sum_{i=2}^{D}\gamma ^{(i)}k^{i}$. In
deriving (\ref{psi2}) the relations
\begin{equation}
\left( \gamma ^{(1)}\partial _{z}-\frac{1}{z}\gamma ^{(0)}\nu \right) K_{\nu
}\left( z\right) =-K_{\nu }\left( z\right) \gamma ^{(1)},  \label{rel3}
\end{equation}%
and $\boldsymbol{\gamma \,}\mathbf{k}K_{\nu }\left( \lambda \rho \right)
=K_{\nu }\left( \lambda \rho \right) \boldsymbol{\gamma \,}\mathbf{k}$ have
been used with $\nu =i\omega -\gamma ^{(0)}\gamma ^{(1)}/2$.

Introducing a new constant spinor $\chi _{\eta }^{(+)}(\mathbf{k})$, the
positive energy fermionic modes are presented in the form%
\begin{equation}
\psi _{\sigma }^{(+)}=N_{\sigma }e^{-i\omega \tau +i\mathbf{k}\cdot \mathbf{x%
}}K_{i\omega -\frac{1}{2}\gamma ^{(0)}\gamma ^{(1)}}\left( \lambda \rho
\right) P(\mathbf{k})\chi _{\eta }^{(+)}(\mathbf{k}),  \label{psip}
\end{equation}%
where $\sigma =(\omega ,\mathbf{k},\eta )$ presents the set of quantum
numbers specifying the modes. Here, $\eta $ enumerates the spinorial degrees
of freedom. For the operator (\ref{Pk}) the properties
\begin{equation}
P^{2}(\mathbf{k})=P(\mathbf{k}),\;P^{\dagger }(\mathbf{k})=P(\mathbf{k}),
\label{Pk1}
\end{equation}%
can be easily checked. We specify the spinors $\chi _{\eta }^{(+)}(\mathbf{k}%
)$ by the relation%
\begin{equation}
P(\mathbf{k})\chi _{\eta }^{(+)}(\mathbf{k})=\chi _{\eta }^{(+)}(\mathbf{k}),
\label{Relxi1}
\end{equation}%
by the orthonormality condition
\begin{equation}
\chi _{\eta }^{(+)\dagger }(\mathbf{k})\chi _{\eta ^{\prime }}^{(+)}(\mathbf{%
k})=\delta _{\eta \eta ^{\prime }},  \label{Relxi2}
\end{equation}%
and by the completeness relation%
\begin{equation}
\sum_{\eta }\chi _{\eta \alpha }^{(+)}(\mathbf{k})\chi _{\eta \beta
}^{(+)\dagger }(\mathbf{k})=P_{\alpha \beta }(\mathbf{k}),  \label{xirel}
\end{equation}%
where $\alpha $ and $\beta $ are spinor indices. The conditions (\ref{Relxi1}%
)-(\ref{xirel}) are the generalizations of the respective relations in \cite%
{Cand78} for a massless field in 4-dimensional spacetime. Note that in \cite%
{Cand78} the Majorana representation was used in which the components of the
spinor $\psi (x)$ and the Dirac matrices are real. In our consideration the
representation is not fixed. With the condition (\ref{Relxi1}), the
following final expression is obtained for the positive energy fermionic
normal modes:%
\begin{equation}
\psi _{\sigma }^{(+)}=N_{\sigma }e^{-i\omega \tau +i\mathbf{k}\cdot \mathbf{x%
}}K_{i\omega -\frac{1}{2}\gamma ^{(0)}\gamma ^{(1)}}\left( \lambda \rho
\right) \chi _{\eta }^{(+)}(\mathbf{k}).  \label{psip2}
\end{equation}%
Note that the energy measured by an observer with $\left( \rho ,\mathbf{x}%
\right) =\mathrm{const}$ is given by $\varepsilon _{\rho }=\omega /\rho $.

The constant $N_{\sigma }$ in (\ref{psip2}) is determined from the
normalization condition $\int d^{D}x\,\sqrt{|g|}\bar{\psi}_{\sigma ^{\prime
}}^{(+)}\gamma ^{0}\psi _{\sigma }^{(+)}=\delta _{\sigma \sigma ^{\prime }}$%
, where $g$ is the determinant of the metric tensor and the Dirac conjugate $%
\bar{\psi}(x)$ for the field $\psi (x)$ is defined as $\bar{\psi}(x)=\psi
^{\dagger }(x)\gamma ^{(0)}$. Here, $\delta _{\sigma \sigma ^{\prime }}$ is
understood as Dirac delta function for continuous quantum numbers and as
Kronecker symbol for discrete ones. By taking into account that $\gamma
^{0}=\gamma ^{(0)}/\rho $ and $\sqrt{|g|}=\rho $, the orthonormality
condition for the modes (\ref{psip2}) is reduced to
\begin{equation}
e^{i\left( \omega ^{\prime }-\omega \right) \tau }\left\vert N_{\sigma
}\right\vert ^{2}\chi _{\eta ^{\prime }}^{(+)\dagger }(\mathbf{k}%
)\int_{0}^{\infty }d\rho \,K_{i\omega ^{\prime }+\frac{1}{2}\gamma
^{(0)}\gamma ^{(1)}}\left( \lambda \rho \right) K_{i\omega -\frac{1}{2}%
\gamma ^{(0)}\gamma ^{(1)}}\left( \lambda \rho \right) \chi _{\eta }^{(+)}(%
\mathbf{k})=\frac{\delta \left( \omega ^{\prime }-\omega \right) }{\left(
2\pi \right) ^{D-1}}\delta _{\eta \eta ^{\prime }}.  \label{nc2}
\end{equation}%
For the evaluation of the integral in the right-hand side we use the result
\cite{Cand78}%
\begin{equation}
\int_{0}^{\infty }d\rho \,K_{i\omega ^{\prime }+\frac{1}{2}\gamma
^{(0)}\gamma ^{(1)}}\left( \lambda \rho \right) K_{i\omega -\frac{1}{2}%
\gamma ^{(0)}\gamma ^{(1)}}\left( \lambda \rho \right) =\frac{\pi ^{2}}{%
2\lambda }\left[ \frac{\delta \left( \omega ^{\prime }-\omega \right) }{%
\cosh \left( \pi \omega \right) }+\frac{i\gamma ^{(0)}\gamma ^{(1)}}{\sinh
\left( \pi \omega ^{\prime }\right) -\sinh \left( \pi \omega \right) }\right]
.  \label{IntKK}
\end{equation}%
The contribution of the second term in the square brackets of (\ref{IntKK})
to the normalization integral in (\ref{nc2}) contains the combination%
\begin{equation}
\chi _{\eta ^{\prime }}^{(+)\dagger }(\mathbf{k})\gamma ^{(0)}\gamma
^{(1)}\chi _{\eta }^{(+)}(\mathbf{k}).  \label{Prodgam}
\end{equation}%
By taking into account $\chi _{\eta ^{\prime }}^{(+)\dagger }(\mathbf{k}%
)=\chi _{\eta ^{\prime }}^{(+)\dagger }(\mathbf{k})P(\mathbf{k})$ and the
relation%
\begin{equation}
\left( 1+i\gamma ^{(1)}\frac{\boldsymbol{\gamma \,}\mathbf{k}+m}{\lambda }%
\right) \chi _{\eta }^{(+)}(\mathbf{k})=0,  \label{Rel0}
\end{equation}%
it can be seen that $\chi _{\eta ^{\prime }}^{(+)\dagger }(\mathbf{k})\gamma
^{(0)}\gamma ^{(1)}\chi _{\eta }^{(+)}(\mathbf{k})=0$. Hence, the second
term in the square brackets of (\ref{IntKK}) does not contribute to the
normalization integral in (\ref{nc2}). As a consequence, for the
normalization coefficient we get%
\begin{equation}
\left\vert N_{\sigma }\right\vert ^{2}=\frac{\lambda \cosh \left( \pi \omega
\right) }{2^{D-2}\pi ^{D+1}},  \label{Nsig}
\end{equation}%
where $\lambda $ is given by (\ref{lam}).

The negative energy fermionic modes are found in a similar way:%
\begin{equation}
\psi _{\sigma }^{(-)}=N_{\sigma }e^{i\omega \tau +i\mathbf{k}\cdot \mathbf{x}%
}K_{i\omega +\frac{1}{2}\gamma ^{(0)}\gamma ^{(1)}}\left( \lambda \rho
\right) \chi _{\eta }^{(-)}(\mathbf{k}),  \label{psim}
\end{equation}%
with the same normalization coefficient. The constant spinors $\chi _{\eta
}^{(-)}(\mathbf{k})$ obey the same relations (\ref{Relxi1})-(\ref{xirel})
with the $N\times N$ matrix $P(\mathbf{k})$ from (\ref{Pk}). It can be
checked that the positive and negative energy modes are orthogonal: $\int
d^{D}x\,\psi _{\sigma ^{\prime }}^{(-)\dagger }\psi _{\sigma }^{(+)}=0$. The
orthogonality integral is reduced to
\begin{equation}
\chi _{\eta ^{\prime }}^{(-)\dagger }(\mathbf{k})\int_{0}^{\infty }d\rho
\,K_{i\omega ^{\prime }-\frac{1}{2}\gamma ^{(0)}\gamma ^{(1)}}\left( \lambda
\rho \right) K_{i\omega -\frac{1}{2}\gamma ^{(0)}\gamma ^{(1)}}\left(
\lambda \rho \right) \chi _{\eta }^{(+)}(\mathbf{k}).  \label{orthint}
\end{equation}%
The integral is evaluated with the help of (\ref{IntKK}), where now the
replacement $\omega ^{\prime }\rightarrow -\omega ^{\prime }$ should be
made. By using the relations $\chi _{\eta ^{\prime }}^{(-)\dagger }(\mathbf{k%
})=\chi _{\eta ^{\prime }}^{(-)\dagger }(\mathbf{k})P(\mathbf{k})$ and (\ref%
{Rel0}), it is seen that $\chi _{\eta ^{\prime }}^{(-)\dagger }(\mathbf{k}%
)\gamma ^{(0)}\gamma ^{(1)}\chi _{\eta }^{(+)}(\mathbf{k})=0$. The remaining
part contains the delta function $\delta \left( \omega ^{\prime }+\omega
\right) $. From the relation (\ref{xirel}) and similar relation for $\chi
_{\eta }^{(-)}(\mathbf{k})$ it follows that%
\begin{equation}
\sum_{\eta }\chi _{\eta }^{(j)\dagger }(\mathbf{k})\chi _{\eta }^{(j)}(%
\mathbf{k})=\frac{N}{2},  \label{xirels}
\end{equation}%
for $j=+,-$. In the discussion below this relation will be used in the
evaluation of the mode sums for the fermion condensate and the VEV of the
energy-momentum tensor.

We have described the fermionic modes in the R region of the Minkowski
spacetime. The modes in the L region have the same structure, whereas the
modes in the F and P regions are obtained by respective analytic
continuations (for the analytic continuation of the modes in 4-dimensional
Rindler spacetime see, e.g., \cite{Alsi06,Ueda21}). Note that in those
regions the line element reads $ds_{\mathrm{F,P}}^{2}=d\rho ^{2}-\rho
^{2}d\tau ^{2}-d\mathbf{x}^{2}$ and the roles of the coordinates $\tau $ and
$\rho $ are reversed: they appear as spatial and time coordinates,
respectively. Here the situation is similar to that for Schwarzschild time
and radial coordinates near black hole horizons. In the F and P regions the
special case of the Kasner metric is realized with flat spacetime geometry.
Alternatively, by the coordinate transformation we can present the line
element in the F and P regions in the form corresponding to the Milne
universe. In that representation the spacetime is foliated by constant
negative curvature spatial sections (see Section \ref{sec:ConfRel}).

Note that, by using the procedure described above, we can also construct a
set of fermionic modes in problems where an additional uniformly
accelerating boundaries (mirrors) are present, for example, in the R region.
For a single boundary and reflecting conditions on the fermionic field, the
R region is divided into two subregions $0<\rho <\rho _{0}$ and $\rho
_{0}<\rho <\infty $, where $1/\rho _{0}$ is the proper acceleration of a
planar mirror. In the region between the Rindler horizon and mirror the
dependence on the coordinate $\rho $ is expressed in terms of the linear
combination of the modified Bessel functions $I_{\nu }(z)$ and $K_{\nu }(z)$
and the relative coefficient in that combination is determined by the
boundary condition. In the region $\rho _{0}<\rho <\infty $ the mode
functions should have the form (\ref{psip2}) and (\ref{psim}), but now the
boundary condition on the mirror will lead to discrete spectrum of the
energy $\omega $. The problems for polarization of the Fulling-Rindler
vacuum by uniformly accelerating boundaries in the cases of scalar and
electromagnetic fields have been discussed in Refs. \cite{Cand77}-\cite%
{Saha06}.

\section{Fermionic condensate}

\label{sec:FC}

In this section we investigate the fermionic condensate (FC) by using the
mode functions from the previous section. It is an important physical
characteristic in models of symmetry breaking and dynamical mass generation
and in the studies of phase transitions. In the most popular model of the
Unruh-DeWitt detector interacting with fermionic fields the interaction
Hamiltonian is proportional to the FC operator evaluated at the location of
the detector (see, e.g., \cite{Lang05,Humm16,Sach17,Taga86}) and the FC is
the central quantity in the interpretation of the response for that type of
detectors.

The FC is defined as the VEV $\left\langle 0\right\vert \bar{\psi}\psi
\left\vert 0\right\rangle =\left\langle \bar{\psi}\psi \right\rangle $,
where $\left\vert 0\right\rangle $ stands for the vacuum state (the
Fulling-Rindler vacuum in the consideration at hand) and $\bar{\psi}=\psi
^{\dagger }\gamma ^{(0)}$. The FC is expressed in terms of the fermionic
Hadamard function $S^{(1)}(x,x^{\prime })$ with the spinorial components%
\begin{equation}
S_{\alpha \beta }^{(1)}(x,x^{\prime })=\left\langle \psi _{\alpha }(x)\bar{%
\psi}_{\beta }(x^{\prime })-\bar{\psi}_{\alpha }(x^{\prime })\psi _{\beta
}(x)\right\rangle ,  \label{S1}
\end{equation}%
where $\alpha $ and $\beta $ are spinorial indices. For the FC one has
\begin{equation}
\left\langle \bar{\psi}\psi \right\rangle =-\frac{1}{2}\lim_{x^{\prime
}\rightarrow x}\mathrm{Tr}\left( S^{(1)}(x,x^{\prime })\right) .  \label{FCF}
\end{equation}%
Expanding the field operator in terms of the complete set of fermionic modes
$\{\psi _{\sigma }^{(+)},\psi _{\sigma }^{(-)}\}$, the following mode sum is
obtained for the trace of the Hadamard function:%
\begin{equation}
\mathrm{Tr}\left( S^{(1)}(x,x^{\prime })\right) =\sum_{\sigma }\sum_{j=+,-}j%
\bar{\psi}_{\sigma }^{(j)}(x^{\prime })\psi _{\sigma }^{(j)}(x),  \label{FC}
\end{equation}%
where%
\begin{equation}
\sum_{\sigma }=\int_{0}^{\infty }d\omega \int d\mathbf{k}\sum_{\eta }.
\label{Sumsig}
\end{equation}

Substituting the mode functions (\ref{psip}) and (\ref{psim}), by using Eq. (%
\ref{rel2}) and the relations
\begin{eqnarray}
\lbrack K_{i\omega -j\frac{1}{2}\gamma ^{(0)}\gamma ^{(1)}}\left( z\right)
]^{\dagger } &=&K_{i\omega +k\frac{1}{2}\gamma ^{(0)}\gamma ^{(1)}}\left(
z\right) ,  \notag \\
\gamma ^{(0)}K_{i\omega -j\frac{1}{2}\gamma ^{(0)}\gamma ^{(1)}}\left(
z\right) &=&K_{i\omega +j\frac{1}{2}\gamma ^{(0)}\gamma ^{(1)}}\left(
z\right) \gamma ^{(0)},  \label{RelKgam}
\end{eqnarray}%
for the products of the normal modes we get%
\begin{equation}
\bar{\psi}_{\sigma }^{(j)}(x^{\prime })\psi _{\sigma }^{(j)}(x)=\left\vert
N_{\sigma }\right\vert ^{2}e^{-ji\omega \Delta \tau +i\mathbf{k}\cdot \Delta
\mathbf{x}}\chi _{\eta }^{(j)\dagger }(\mathbf{k})K_{i\omega +j\frac{1}{2}%
\gamma ^{(0)}\gamma ^{(1)}}\left( \lambda \rho ^{\prime }\right) K_{i\omega
+j\frac{1}{2}\gamma ^{(0)}\gamma ^{(1)}}\left( \lambda \rho \right) \gamma
^{(0)}\chi _{\eta }^{(j)}(\mathbf{k}),  \label{psipm}
\end{equation}%
where $\Delta \tau =\tau -\tau ^{\prime }$ and $\Delta \mathbf{x}=\mathbf{x}-%
\mathbf{x}^{\prime }$. Now, by using the relation (\ref{rel2}) for the
modified Bessel functions, the trace is presented as%
\begin{eqnarray}
\mathrm{Tr}\left( S^{(1)}(x,x^{\prime })\right) &=&\int d\mathbf{k}\,\lambda
e^{i\mathbf{k}\cdot \Delta \mathbf{x}}\int_{0}^{\infty }d\omega \,\frac{%
\cosh \left( \pi \omega \right) }{2^{D-1}\pi ^{D+1}}\sum_{j=+,-}je^{-ji%
\omega \Delta \tau }  \notag \\
&&\times \sum_{\varkappa =\pm 1}K_{i\omega +j\frac{\varkappa }{2}}\left(
\lambda \rho \right) K_{i\omega +j\frac{\varkappa }{2}}\left( \lambda \rho
^{\prime }\right) \sum_{\eta }\chi _{\eta }^{(j)\dagger }(\mathbf{k})\left(
\gamma ^{(0)}-\varkappa \gamma ^{(1)}\right) \chi _{\eta }^{(j)}(\mathbf{k}).
\label{psipm2}
\end{eqnarray}%
The summation over $\eta $ is done by using the result (\ref{xirels}):%
\begin{equation}
\sum_{\eta }\chi _{\eta }^{(j)\dagger }(\mathbf{k})\left( \gamma
^{(0)}-\varkappa \gamma ^{(1)}\right) \chi _{\eta }^{(j)}(\mathbf{k})=%
\mathrm{Tr}\left[ P(\mathbf{k})\left( \gamma ^{(0)}-\varkappa \gamma
^{(1)}\right) \right] =-\frac{\varkappa imN}{2\lambda },  \label{Sumeta}
\end{equation}%
where the relations $\mathrm{Tr}\left[ \gamma ^{(i_{1})}\gamma
^{(i_{2})}\cdots \gamma ^{(i_{2n+1})}\right] =0$ and $\mathrm{Tr}\left[
\gamma ^{(a)}\gamma ^{(b)}\right] =(D+1)\eta ^{ab}$ have been used. This
leads to the result%
\begin{eqnarray}
\mathrm{Tr}\left( S^{(1)}(x,x^{\prime })\right) &=&-\frac{Nm}{2^{D-2}\pi
^{D+1}}\int d\mathbf{k\,}e^{i\mathbf{k}\cdot \Delta \mathbf{x}%
}\int_{0}^{\infty }d\omega \,\cosh \left( \pi \omega \right)  \notag \\
&&\times \cos \left( \omega \Delta \tau \right) \mathrm{Im}\left[ K_{\frac{1%
}{2}-i\omega }\left( \lambda \rho \right) K_{\frac{1}{2}-i\omega }\left(
\lambda \rho ^{\prime }\right) \right] .  \label{FC2}
\end{eqnarray}

The limit in the right-hand side of (\ref{FCF}) is divergent and the
renormalization of the FC is required. The background spacetime is flat and
the divergences are the same as those for the Minkowski vacuum.
Renormalizing the FC in the Minkowski vacuum to zero, for the renormalized
FC in the Fulling-Rindler vacuum we get%
\begin{eqnarray}
\left\langle \bar{\psi}\psi \right\rangle _{\mathrm{FR}} &=&\left\langle
\bar{\psi}\psi \right\rangle -\left\langle \bar{\psi}\psi \right\rangle _{%
\mathrm{M}}  \notag \\
&=&-\frac{1}{2}\lim_{x^{\prime }\rightarrow x}\left[ \mathrm{Tr}\left(
S^{(1)}(x,x^{\prime })\right) -\mathrm{Tr}(S_{\mathrm{MR}}^{(1)}(x,x^{\prime
}))\right] ,  \label{FCren}
\end{eqnarray}%
where $S_{\mathrm{MR}}^{(1)}(x,x^{\prime })$ is the Hadamard function for
the Minkowski vacuum transformed to the Rindler coordinates. The expression
for the latter, adapted for the subtraction in (\ref{FCren}), is provided in
Appendix \ref{sec:AppMink} (see (\ref{Str2})). Plugging the representations (%
\ref{FC2}) and (\ref{Str2}) (the latter multiplied by $\cosh \left( \Delta
\tau /2\right) $) and integrating over the angular coordinates of $\mathbf{k}
$, for the renormalized FC in the Fulling-Rindler vacuum one finds%
\begin{equation}
\left\langle \bar{\psi}\psi \right\rangle _{\mathrm{FR}}=\frac{2^{1-D}mN}{%
\pi ^{\frac{D+3}{2}}\Gamma \left( \frac{D-1}{2}\right) }\int_{0}^{\infty
}d\omega \,e^{-\pi \omega }\int_{m}^{\infty }d\lambda \mathbf{\,}\lambda
\left( \lambda ^{2}-m^{2}\right) ^{\frac{D-3}{2}}\,\mathrm{Im}\left[
K_{1/2-i\omega }^{2}\left( \lambda \rho \right) \right] ,  \label{FCren1}
\end{equation}%
where $D\geq 2$. In the special case $D=1$, with $N=2$ for the irreducible
representation of the Clifford algebra, the corresponding expression has the
form%
\begin{equation}
\left\langle \bar{\psi}\psi \right\rangle _{\mathrm{FR}}=\frac{m}{\pi ^{2}}%
\int_{0}^{\infty }d\omega e^{-\pi \omega }\mathrm{Im}\left[ K_{\frac{1}{2}%
-i\omega }^{2}\left( m\rho \right) \right] .  \label{FCren1D1}
\end{equation}%
For a massless field the FC vanishes. Note that we could write a formal
expression obtained from (\ref{FC2}) taking the coincidence limit in the
integrand:%
\begin{equation}
\left\langle \bar{\psi}\psi \right\rangle =\frac{Nm}{2^{D-1}\pi ^{D+1}}\int d%
\mathbf{k\,}\int_{0}^{\infty }d\omega \cosh \left( \pi \omega \right)
\mathrm{Im}\left[ K_{\frac{1}{2}-i\omega }^{2}\left( \lambda \rho \right) %
\right] .  \label{FC2b}
\end{equation}%
The divergence comes from the term $e^{\pi \omega }/2$ in the definition of
the hyperbolic cosine function. Comparing with (\ref{FCren1D1}) we see that
the subtraction of the Minkowskian VEV is equivalent to the subtraction of
the part with that term.

The ratio $\left\langle \bar{\psi}\psi \right\rangle _{\mathrm{FR}}/m^{D}$
is a function of a single variable $m\rho $. Let us consider the behavior of
that function in asymptotic regions for $D\geq 2$. For $m\rho \gg 1$ the
argument of the modified Bessel function is large and we use the leading
order estimate%
\begin{equation}
\mathrm{Im}\left[ K_{1/2-i\omega }^{2}\left( z\right) \right] \sim -\frac{%
\pi \omega }{2z^{2}}e^{-2z}.  \label{Kas1}
\end{equation}%
This gives%
\begin{equation}
\left\langle \bar{\psi}\psi \right\rangle _{\mathrm{FR}}\approx -\frac{%
Nm^{D}e^{-2m\rho }}{2^{D+1}\pi ^{\frac{D+5}{2}}\left( m\rho \right) ^{\frac{%
D+3}{2}}},\;m\rho \gg 1,  \label{FRas1}
\end{equation}%
and the FC is exponentially small. In particular, for an observer with a
given acceleration the FC is exponentially small for large values of the
mass.

In the opposite limit $m\rho \ll 1$ we introduce a new integration variable $%
u=\lambda \rho $ and put in the integral $m\rho =0$ for the leading order
term:
\begin{equation}
\left\langle \bar{\psi}\psi \right\rangle _{\mathrm{FR}}\approx \frac{%
Nm\left( 2\rho \right) ^{1-D}}{\pi ^{\frac{D+3}{2}}\Gamma \left( \frac{D-1}{2%
}\right) }\int_{0}^{\infty }d\omega \,e^{-\pi \omega }\int_{0}^{\infty }du%
\mathbf{\,}u^{D-2}\,\mathrm{Im}\left[ K_{1/2-i\omega }^{2}\left( u\right) %
\right] .  \label{FRas2}
\end{equation}%
The integral is evaluated by using the formula%
\begin{equation}
\int_{0}^{\infty }du\mathbf{\,}u^{\nu }\mathbf{\,}\,\mathrm{Im}\left[
K_{1/2-i\omega }^{2}\left( u\right) \right] =-\frac{\sqrt{\pi }\Gamma \left(
(\nu +1)/2\right) }{4\Gamma \left( \nu /2+1\right) }\omega \left\vert \Gamma
\left( \frac{\nu }{2}+i\omega \right) \right\vert ^{2},  \label{IntK2}
\end{equation}%
with $\nu =D-2$, and we get%
\begin{equation}
\left\langle \bar{\psi}\psi \right\rangle _{\mathrm{FR}}\approx -\frac{%
Nm\rho ^{1-D}}{2^{D+1}\pi ^{\frac{D}{2}+1}\Gamma \left( \frac{D}{2}\right) }%
\int_{0}^{\infty }d\omega \,\omega e^{-\pi \omega }\left\vert \Gamma \left(
\frac{D}{2}-1+i\omega \right) \right\vert ^{2}.  \label{FRas3}
\end{equation}%
Note that for the gamma function in the integrand we have%
\begin{equation}
\left\vert \Gamma \left( \frac{n}{2}+i\omega \right) \right\vert ^{2}=\frac{%
2\pi \omega ^{n-1}e^{\pi \omega }B_{n}(\omega )}{e^{2\pi \omega }-\left(
-1\right) ^{n}},  \label{Gam}
\end{equation}%
where
\begin{equation}
B_{n}(\omega )=\prod\limits_{l=1}^{l_{n}}\left[ 1+\left( \frac{l-\{n/2\}}{%
\omega }\right) ^{2}\right] ,  \label{Bn}
\end{equation}%
for $n\geq 2$ and $B_{0}=B_{1}=1$. In (\ref{Bn}), $\{n/2\}$ is the
fractional part of $n/2$ and we have defined%
\begin{equation}
l_{n}=n/2-1+\{n/2\}.  \label{ln}
\end{equation}%
Note that, for a given value of the mass the condition $m\rho \ll 1$
corresponds to points near the Rindler horizon $\rho =0$. From (\ref{FRas3})
we conclude that near the Rindler horizon the FC behaves like $1/\rho ^{D-1}$%
.

\section{VEV of the energy-momentum tensor}

\label{sec:EMT}

We turn to the study of another important local characteristic of the
fermionic vacuum, namely, the expectation value of the energy-momentum
tensor. In addition to describing the local properties of the vacuum state,
it appears as a source of gravity in the semiclassical Einstein equations
and determines the back reaction of quantum effects on the spacetime
geometry. The possibility to measure the VEV of the energy-momentum tensor
for a scalar field in non-gravitational way by using particle detectors has
been discussed in \cite{Ford93}. Similar to the case of the FC, the point
splitting procedure will be used for regularization. The corresponding mode
sum is expressed as
\begin{equation}
\langle T_{\mu \nu }\rangle =-\frac{i}{4}\lim_{x^{\prime }\rightarrow
x}\sum_{\sigma }\sum_{j=+,-}j[\bar{\psi}_{\sigma }^{(j)}(x^{\prime })\gamma
_{(\mu }\nabla _{\nu )}\psi _{\sigma }^{(j)}(x)-(\nabla _{(\mu }^{\prime }%
\bar{\psi}_{\sigma }^{(j)}(x^{\prime }))\gamma _{\nu )}\psi _{\sigma
}^{(j)}(x)]\ ,  \label{modesum}
\end{equation}%
where the brackets denote symmetrization with respect to the enclosed
indices and the action of the covariant derivative operator on the Dirac
conjugate spinor is given by $\nabla _{\mu }\bar{\psi}_{\sigma
}^{(j)}=\partial _{\mu }\bar{\psi}_{\sigma }^{(j)}-\bar{\psi}_{\sigma
}^{(j)}\Gamma _{\mu }$ with the spin connection from (\ref{Gamu}). The
summation $\sum_{\sigma }$ is understood in the sense (\ref{Sumsig}).

The evaluation procedure for the energy-momentum tensor is similar to that
for the FC and we will omit the details. Substituting the mode functions (%
\ref{psip2}), (\ref{psim}) and using the relation (\ref{rel2}), for the
diagonal components we get (no summation over $\mu $)%
\begin{equation}
\langle T_{\mu }^{\mu }\rangle =-\frac{2^{1-D}N}{\pi ^{D+1}}\lim_{x^{\prime
}\rightarrow x}\int d\mathbf{k\,}e^{i\mathbf{k}\cdot \Delta \mathbf{x}%
}\int_{0}^{\infty }d\omega \,\cosh \left( \pi \omega \right) \cos \left(
\omega \Delta \tau \right) F_{\omega ,\lambda }^{(\mu )}(\rho ,\rho ^{\prime
}),  \label{Tmu}
\end{equation}%
with the functions%
\begin{eqnarray}
F_{\omega ,\lambda }^{(0)}(\rho ,\rho ^{\prime }) &=&\frac{\omega \lambda }{%
\rho }\,\mathrm{Re}\left[ K_{\frac{1}{2}-i\omega }\left( \lambda \rho
\right) K_{\frac{1}{2}+i\omega }\left( \lambda \rho ^{\prime }\right) \right]
,  \notag \\
F_{\omega ,\lambda }^{(1)}(\rho ,\rho ^{\prime }) &=&\frac{\lambda ^{2}}{2}\,%
\mathrm{Im}\left[ K_{\frac{1}{2}-i\omega }\left( \lambda \rho \right) K_{%
\frac{1}{2}+i\omega }^{\prime }\left( \lambda \rho ^{\prime }\right) +K_{%
\frac{1}{2}+i\omega }^{\prime }\left( \lambda \rho \right) K_{\frac{1}{2}%
-i\omega }\left( \lambda \rho ^{\prime }\right) \right] ,  \notag \\
F_{\omega ,\lambda }^{(l)}(\rho ,\rho ^{\prime }) &=&\frac{k^{2}}{D-1}\,%
\mathrm{Im}\left[ K_{\frac{1}{2}-i\omega }\left( \lambda \rho ^{\prime
}\right) K_{\frac{1}{2}-i\omega }\left( \lambda \rho \right) \right] ,
\label{Fmu}
\end{eqnarray}%
where in the last expression $l=2,\ldots ,D$. The off-diagonal components
vanish. Again, the limit in the right-hand side of (\ref{Tmu}) is divergent
and in order to obtain the renormalized VEVs we need to subtract the
corresponding expectation values for the Minkowski vacuum. Similar to the
case of the FC it can be seen that the subtraction is equivalent to omitting
the term $e^{\pi \omega }/2$ in the expression for the hyperbolic cosine
function $\cosh \left( \pi \omega \right) $. In this way, for the
renormalized VEV, $\langle T_{\mu }^{\nu }\rangle _{\mathrm{FR}}=\langle
T_{\mu }^{\nu }\rangle -\langle T_{\mu }^{\nu }\rangle _{\mathrm{M}}$, we
get (no summation over $\mu $)%
\begin{equation}
\langle T_{\mu }^{\mu }\rangle _{\mathrm{FR}}=-\frac{2^{1-D}N}{\pi ^{\frac{%
D+3}{2}}\Gamma \left( \frac{D-1}{2}\right) }\int_{0}^{\infty }d\omega
\,e^{-\pi \omega }\int_{m}^{\infty }d\lambda \,\mathbf{\,}\lambda \left(
\lambda ^{2}-m^{2}\right) ^{\frac{D-3}{2}}F_{\omega ,\lambda }^{(\mu )}(\rho
),  \label{Tmu2}
\end{equation}%
where $D\geq 2$ and the functions for separate components are defined as%
\begin{eqnarray}
F_{\omega ,\lambda }^{(0)}(\rho ) &=&\frac{\omega \lambda }{\rho }\left\vert
K_{\frac{1}{2}-i\omega }\left( \lambda \rho \right) \right\vert ^{2},  \notag
\\
F_{\omega ,\lambda }^{(1)}(\rho ) &=&-\lambda ^{2}\,\mathrm{Im}\left[ K_{%
\frac{1}{2}+i\omega }\left( \lambda \rho \right) K_{\frac{1}{2}-i\omega
}^{\prime }\left( \lambda \rho \right) \right] ,  \notag \\
F_{\omega ,\lambda }^{(l)}(\rho ) &=&\frac{\lambda ^{2}-m^{2}}{D-1}\,\mathrm{%
Im}\left[ K_{\frac{1}{2}-i\omega }^{2}\left( \lambda \rho \right) \right] ,
\label{Fmuc}
\end{eqnarray}%
with $l=2,\ldots ,D$. In the special case $D=1$ one obtains%
\begin{eqnarray}
\langle T_{0}^{0}\rangle _{\mathrm{FR}} &=&-\frac{m}{\pi ^{2}\rho }%
\int_{0}^{\infty }d\omega \,\omega e^{-\pi \omega }\,\left\vert K_{\frac{1}{2%
}-i\omega }\left( m\rho \right) \right\vert ^{2},  \notag \\
\langle T_{1}^{1}\rangle _{\mathrm{FR}} &=&-\frac{m^{2}}{\pi ^{2}}%
\int_{0}^{\infty }d\omega \,e^{-\pi \omega }\,\mathrm{Im}\left[ K_{\frac{1}{2%
}-i\omega }\left( m\rho \right) K_{\frac{1}{2}+i\omega }^{\prime }\left(
m\rho \right) \right] .  \label{TmuD1}
\end{eqnarray}%
The off-diagonal components of the renormalized mean energy-momentum tensor
are zero. Alternative representations of the FC and mean energy-momentum
tensor will be provided below in Section \ref{sec:AlterRep}. In particular,
they show that (no summation over $l$) $\langle T_{l}^{l}\rangle _{\mathrm{FR%
}}=\langle T_{1}^{1}\rangle _{\mathrm{FR}}$ for $l=2,3,\ldots ,D$, and the
vacuum stresses are isotropic. Note that this property does not follow from
the symmetry of the problem. For example, in the case of a scalar field the
stresses $\langle T_{1}^{1}\rangle _{\mathrm{FR}}$ and $\langle
T_{2}^{2}\rangle _{\mathrm{FR}}$, in general, differ (see \cite{Saha02}).
They coincide for a conformally coupled massless field only (see below). In
analogy of the perfect fluid, we can interpret the quantities (no summation
over $l$) $-\langle T_{l}^{l}\rangle _{\mathrm{FR}}$ as effective pressures
along the respective directions.

The energy density corresponding to Eq. (\ref{Tmu2}) is always negative, $%
\langle T_{\mu }^{\mu }\rangle _{\mathrm{FR}}<0$. The representations given
in Section \ref{sec:AlterRep} explicitly show that the vacuum pressures $%
-\langle T_{l}^{l}\rangle _{\mathrm{FR}}$ are negative as well. We expect
that the renormalized VEVs will obey the covariant conservation equation $%
\nabla _{\mu }\langle T_{\nu }^{\mu }\rangle _{\mathrm{FR}}=0$. For the
geometry at hand it is reduced to the relation $\langle T_{0}^{0}\rangle _{%
\mathrm{FR}}=\partial _{\rho }\left( \rho \langle T_{1}^{1}\rangle _{\mathrm{%
FR}}\right) $. By using the differential equation for the modified Bessel
functions it is easy to check that this relation indeed takes place. Writing
the VEV (\ref{Tmu2}) in the form $\langle T_{\mu }^{\nu }\rangle _{\mathrm{M}%
}-\langle T_{\mu }^{\nu }\rangle =-\langle T_{\mu }^{\nu }\rangle _{\mathrm{%
FR}}$, we can interpret the tensor $-\langle T_{\mu }^{\nu }\rangle _{%
\mathrm{FR}}$ as the energy-momentum tensor of the Minkowski vacuum with
respect to the Fulling-Rindler vacuum state. In this interpretation the
respective energy density is positive.

For a massless field the integrals over $\lambda $ in (\ref{Tmu2}) for $\mu
\neq 1$ are evaluated by the formula (\ref{IntK2}). The integral for $\mu =1$
is obtained from the general formula in \cite{Prud2} for the product of
modified Bessel functions:%
\begin{equation}
\int_{0}^{\infty }dx\mathbf{\,}x^{D}\,\mathrm{Im}\left[ K_{1/2+i\omega
}\left( x\right) K_{1/2-i\omega }^{\prime }\left( x\right) \right] =\frac{%
\sqrt{\pi }\Gamma \left( \frac{D-1}{2}\right) }{4D\Gamma \left( D/2\right) }%
\omega \left\vert \Gamma \left( \frac{D}{2}+i\omega \right) \right\vert ^{2}.
\label{IntKKd}
\end{equation}%
For the renormalized VEV the following result is obtained%
\begin{equation}
\langle T_{\mu }^{\nu }\rangle _{\mathrm{FR}}=\frac{N\left( 2\rho \right)
^{-D-1}}{\pi ^{\frac{D}{2}+1}\Gamma \left( \frac{D}{2}\right) }%
\int_{0}^{\infty }d\omega \,\omega e^{-\pi \omega }\left\vert \Gamma \left(
\frac{D}{2}+i\omega \right) \right\vert ^{2}\mathrm{diag}\left( -1,\frac{1}{D%
},\ldots ,\frac{1}{D}\right) .  \label{Tmum0}
\end{equation}%
The corresponding stresses are isotropic and the energy-momentum tensor is
traceless. We could expect the latter property on the basis of conformal
invariance of a massless fermionic field and flatness of the background
geometry (zero trace anomaly).

An equivalent representation of the mean energy-momentum tensor for a
massless field is obtained by using the relation (\ref{Gam}). This gives%
\begin{equation}
\langle T_{\mu }^{\nu }\rangle _{\mathrm{FR}}=\frac{A_{D}^{\mathrm{(ferm)}}}{%
\rho ^{D+1}}\mathrm{diag}\left( -1,\frac{1}{D},\ldots ,\frac{1}{D}\right) ,
\label{Tmum0b}
\end{equation}%
where%
\begin{equation}
A_{D}^{\mathrm{(ferm)}}=\frac{2^{-D}N}{\pi ^{\frac{D}{2}}\Gamma \left( \frac{%
D}{2}\right) }\int_{0}^{\infty }d\omega \,\frac{\omega ^{D}B_{D}(\omega )}{%
e^{2\pi \omega }-\left( -1\right) ^{D}}.  \label{ADf}
\end{equation}%
Here, $B_{D}(\omega )$ is defined by (\ref{Bn}) for $D\geq 2$ and $%
B_{0}=B_{1}=1$. In the special case $D=3$ we get%
\begin{equation}
\langle T_{\mu }^{\nu }\rangle _{\mathrm{FR}}=\frac{1}{\pi ^{2}\rho ^{4}}%
\int_{0}^{\infty }d\omega \,\frac{\omega \left( \omega ^{2}+\frac{1}{4}%
\right) }{e^{2\pi \omega }+1}\mathrm{diag}\left( -1,\frac{1}{3},\ldots ,%
\frac{1}{3}\right) ,  \label{Tmum0D3}
\end{equation}%
which coincides with the result from \cite{Cand78}. We can write the
integral in (\ref{Tmum0D3}) in terms of the proper energy $\varepsilon
_{\rho }=\omega /\rho $, measured by an observer with proper acceleration $%
1/\rho $. The respective factor $\left( e^{2\pi \rho \varepsilon _{\rho
}}+1\right) ^{-1}$ is interpreted as an indication of the thermal nature of
inertial vacuum with respect to a uniformly accelerating observer \cite%
{Full73}-\cite{Cand78}. The corresponding temperature (Unruh temperature) is
given by $T=1/(2\pi \rho )$. An interesting point to be mentioned is that in
even number of spatial dimensions the thermal factor for Dirac field is of
bosonic type, $(e^{2\pi \rho \varepsilon _{\rho }}-1)^{-1}$. Similar
features in the response of a uniformly accelerating Unruh-DeWitt detector
interacting with Dirac field prepared in the Minkowski vacuum have been
observed in \cite{Taga85,Taga86}. Note that for a scalar field in general
number of spatial dimension the thermal factor is in the form $[e^{2\pi
\omega }+(-1)^{D}]^{-1}$ (see \cite{Taga86,Saha02,Oogu86}). The appearance
of the Fermi-Dirac type factor for scalar fields in even number of spatial
dimensions has been further discussed in \cite{Unru86,Srir02} (see also \cite%
{Arre21} for more recent consideration and references). The calculation of
Rindler noise on the basis of the fluctuation-dissipation theorem, that
gives a hint to the origin of inversion of statistics, is presented in \cite%
{Tera99}. The coefficient $A_{D}^{\mathrm{(ferm)}}$ in (\ref{Tmum0b}) is
expressed in terms of the products of the Euler gamma function and the
Riemann zeta function $\zeta (z)$. For example, $A_{3}^{\mathrm{(ferm)}%
}=17/(1920\pi ^{2})$ and $A_{4}^{\mathrm{(ferm)}}=[\pi ^{2}\zeta (3)+3\zeta
(5)]/\left( 16\pi ^{7}\right) $. In Table \ref{Table1} we give the numerical
values of the coefficient for $D=2,\ldots ,10$.

\begin{table}[tbph]
\begin{center}
\begin{tabular}{|c|c|c|c|c|c|c|c|c|c|}
\hline
$D$ & 2 & 3 & 4 & 5 & 6 & 7 & 8 & 9 & 10 \\ \hline
$10^4\times A_{D}^{\mathrm{(ferm)}}$ & 15.43 & 8.97 & 3.10 & 2.45 & 1.08 &
1.03 & 0.54 & 0.60 & 0.35 \\ \hline
\end{tabular}%
\end{center}
\caption{The values of the coefficient in the expression (\protect\ref%
{Tmum0b}) for the vacuum energy-momentum tensor in different spatial
dimensions.}
\label{Table1}
\end{table}

As seen from (\ref{Tmu2}), for a massive field the ratio $\langle T_{\mu
}^{\nu }\rangle _{\mathrm{FR}}/m^{D+1}$ depends on the mass and coordinate $%
\rho $ in the form of dimensionless combination $m\rho $. Let us consider
the behavior of the VEV in asymptotic regions of that combination. For $%
m\rho \gg 1$ the argument of the modified Bessel functions in (\ref{Tmu2}), (%
\ref{Fmuc}) is large and the dominant contribution to the integral over $%
\lambda $ comes from the region near the lower limit of the integration. By
using the asymptotic (\ref{Kas1}) and%
\begin{equation}
\mathrm{Im}\left[ K_{\frac{1}{2}+i\omega }\left( z\right) K_{\frac{1}{2}%
-i\omega }^{\prime }\left( z\right) \right] \sim \frac{\pi \omega }{4z^{3}}%
e^{-2z},  \label{Kas2}
\end{equation}%
in the leading order we get (no summation over $l$)%
\begin{equation}
\langle T_{0}^{0}\rangle _{\mathrm{FR}}\approx -\frac{\langle
T_{l}^{l}\rangle _{\mathrm{FR}}}{2m\rho }\approx -\frac{Nm^{D+1}e^{-2m\rho }%
}{2^{D+1}\pi ^{\frac{D+5}{2}}\left( m\rho \right) ^{\frac{D+3}{2}}},
\label{T00as1}
\end{equation}%
for $l=1,2,\ldots ,D$. In particular, for large masses of the field quanta
the vacuum energy-momentum tensor is exponentially small. An interesting
feature seen from the asymptotic estimate (\ref{T00as1}) is that the
absolute value of the energy density is much smaller than the absolute value
of the pressure, $|\langle T_{0}^{0}\rangle _{\mathrm{FR}}|\ll |\langle
T_{1}^{1}\rangle _{\mathrm{FR}}|$. For classical sources $T_{0}^{0}\geq
|T_{1}^{1}|$ and in the non relativistic limit $|T_{1}^{1}|\ll T_{0}^{0}$.
In the opposite limit $m\rho \ll 1$, the leading term in the expansion over $%
m\rho $ coincides with the VEV for a massless field given by (\ref{Tmum0b}).
In particular, for a given mass, near the Rindler horizon the diagonal
components of the energy-momentum tensor behave as $1/\rho ^{D+1}$.

It is of interest to compare the VEV (\ref{Tmum0b}) for a massless Dirac
field with the respective result for a massless scalar field with curvature
coupling parameter $\xi $. The corresponding mean energy-momentum tensor is
given by the formula (no summation over $\mu $) \cite{Saha02}%
\begin{equation}
\langle T_{\mu }^{\nu }\rangle _{\mathrm{FR}}^{\mathrm{(sc)}}=-\frac{\delta
_{\mu }^{\nu }\rho ^{-D-1}}{2^{D-1}\pi ^{\frac{D}{2}}\Gamma \left( \frac{D}{2%
}\right) }\int_{0}^{\infty }d\omega \,\frac{\omega ^{D}B_{D}^{\mathrm{(sc)}%
}(\omega )f_{0}^{(\mu )}\left( \omega \right) }{e^{2\pi \omega }+\left(
-1\right) ^{D}},  \label{Tmusc}
\end{equation}%
with the functions%
\begin{eqnarray}
f_{0}^{(0)}\left( \omega \right) &=&-Df_{0}^{(1)}\left( \omega \right)
=1+D(D-1)\frac{\xi _{D}-\xi }{\omega ^{2}},  \notag \\
f_{0}^{(l)}\left( \omega \right) &=&-\frac{1}{D}+(D-1)^{2}\frac{\xi _{D}-\xi
}{\omega ^{2}},  \label{f00}
\end{eqnarray}%
for $l=2,3,\ldots ,D$. Here, $\xi _{D}=(D-1)/(4D)$ and the factor $B_{D}^{%
\mathrm{(sc)}}(\omega )$ is the scalar analog of (\ref{Bn}), defined as%
\begin{equation}
B_{n}^{\mathrm{(sc)}}(\omega )=\prod\limits_{l=1}^{l_{n}}\left[ 1+\left(
\frac{l-\frac{1}{2}-\{\frac{n}{2}\}}{\omega }\right) ^{2}\right] ,
\label{BDsc}
\end{equation}%
for $D\geq 3$ and $B_{1}^{\mathrm{(sc)}}(\omega )=B_{2}^{\mathrm{(sc)}%
}(\omega )=1$. As it has been already mentioned, for scalar field, in
general, the stresses are anisotropic. For a conformally coupled field $\xi
=\xi _{D}$ and the formula (\ref{Tmusc}) is reduced to
\begin{equation}
\langle T_{\mu }^{\nu }\rangle _{\mathrm{FR}}^{\mathrm{(sc)}}=\frac{A_{D}^{%
\mathrm{(sc)}}}{\rho ^{D+1}}\mathrm{diag}\left( -1,\frac{1}{D},\ldots ,\frac{%
1}{D}\right) ,  \label{Tmusc2}
\end{equation}%
with the notation%
\begin{equation}
A_{D}^{\mathrm{(sc)}}=\frac{2^{1-D}}{\pi ^{\frac{D}{2}}\Gamma \left( \frac{D%
}{2}\right) }\int_{0}^{\infty }d\omega \,\frac{\omega ^{D}B_{D}^{\mathrm{(sc)%
}}(\omega )}{e^{2\pi \omega }+\left( -1\right) ^{D}}.  \label{ADs}
\end{equation}%
This result generalizes the relation between the VEVs for massless scalar
and Dirac fields, conjectured in \cite{Scia81} (see also \cite{Birr82,Cand78}%
).

For even values of spatial dimension $D$ the Clifford algebra of Dirac
matrices has two inequivalent irreducible representations. These two
representations can be constructed by using an irreducible representation $%
\gamma ^{(a)}$, $a=0,1,\ldots ,D-1$, for $(D-1)$-dimensional space and
adding the matrix $\gamma _{(s)}^{(D)}=i^{2\{D/4\}}s\gamma ^{(D)}$, where $%
\gamma ^{(D)}=\gamma ^{(0)}\gamma ^{(1)}\cdots \gamma ^{(D-1)}$ and $s=\pm 1$
(see, e.g., \cite{Shim85}). Two inequivalent representations for even $D$
are realized by two sets $\gamma _{(s)}^{(a)}=\{\gamma ^{(0)},\cdots \gamma
^{(D-1)},\gamma _{(s)}^{(D)}\}$ with $s=\pm 1$. The consideration we have
presented above does not depend on the representation of Dirac matrices.
Hence, we conclude that the FC and the mean energy-momentum tensor coincide
for Dirac fields $\psi _{(s)}$ realizing two inequivalent representations of
the Clifford algebra in even numbers of spatial dimensions. Note that for
even $D$ the mass term in the Lagrangian density for a field $\psi _{(s)}$
with given $s$ is not invariant under the parity (P) transformation.
Additionally, the mass term is not invariant under the charge conjugation
(C) for $\{D/4\}=0$ and under the time reversal (T) for $\{D/4\}=1/2$.
Fermionic models invariant under those transformations can be constructed
combining two fields $\psi _{(+1)}$ and $\psi _{(-1)}$ with the same masses.
In those models the VEVs are obtained from the expressions presented above
with an additional coefficient 2.

\section{Alternative representations and numerical analysis}

\label{sec:AlterRep}

In this section, equivalent representations are provided for the VEVs, more
convenient in numerical evaluations. We start with the FC given by (\ref%
{FCren1}). From the integral representation (\ref{RepKK1}) for the product
of the modified Bessel functions it follows that
\begin{equation}
\mathrm{Im}\left[ K_{1/2-i\omega }^{2}\left( \lambda \rho \right) \right]
=\int_{-\infty }^{+\infty }du\,e^{-u}\sin \left( 2\omega u\right)
K_{0}\left( 2\lambda \rho \cosh u\right) .  \label{IntKK2}
\end{equation}%
Substituting this in (\ref{FCren1}), the integral over $\lambda $ is
evaluated by using the formula \cite{Prud2}%
\begin{equation}
\int_{m}^{\infty }d\lambda \,\mathbf{\,}\lambda ^{\nu +1}\left( \lambda
^{2}-m^{2}\right) ^{\beta -1}K_{\nu }\left( c\lambda \right) =2^{\beta
-1}m^{2\left( \beta +\nu \right) }\Gamma \left( \beta \right) c^{\nu
}f_{\beta +\nu }\left( mc\right) ,  \label{IntK}
\end{equation}%
where we have introduced the function%
\begin{equation}
f_{\nu }\left( z\right) =z^{-\nu }K_{\nu }\left( z\right) .  \label{fnu}
\end{equation}%
The integration over $\omega $ is elementary and we get%
\begin{equation}
\left\langle \bar{\psi}\psi \right\rangle _{\mathrm{FR}}=-\frac{2Nm^{D}}{%
\left( 2\pi \right) ^{\frac{D+3}{2}}}\int_{0}^{\infty }du\,\frac{u\sinh u}{%
u^{2}+\pi ^{2}/4}f_{\frac{D-1}{2}}\left( 2m\rho \cosh u\right) .
\label{FCalt}
\end{equation}%
This shows that FC is always negative. Note that this formula could also be
directly obtained from the representation (\ref{S16}) for the trace of the
Hadamard function derived in Appendix \ref{sec:App3}.

\begin{figure}[tbph]
\begin{center}
\epsfig{figure=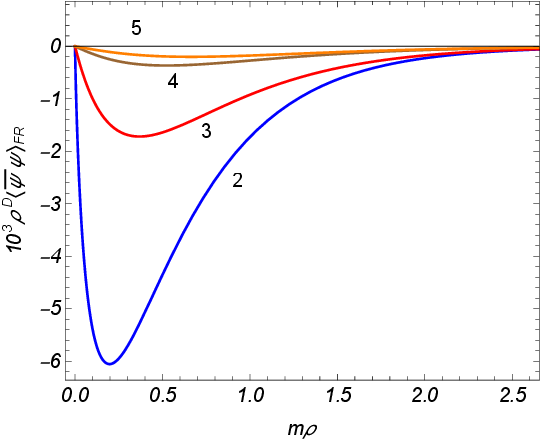,width=7.5cm,height=6.5cm}
\end{center}
\caption{The FC in units of $1/\protect\rho ^{D}$ versus the dimensionless
combination $m\protect\rho $ in spatial dimensions $D=2,3,4,5$ (the numbers
near the curves).}
\label{figFC}
\end{figure}

The FC is monotonically decreasing function of the proper acceleration $%
1/\rho $. It is exponentially suppressed by the factor $e^{-2m\rho }$ for
large values of $\rho $ and behaves like $1/\rho ^{D-1}$ for small $\rho $
(near the Rindler horizon). In Figure \ref{figFC} we have plotted the
quantity $\rho ^{D}\left\langle \bar{\psi}\psi \right\rangle _{\mathrm{FR}}$
as a function of dimensionless combination $m\rho $ for different values of
the spatial dimension. For a massless field the FC vanishes.

\begin{figure}[tbph]
\begin{center}
\epsfig{figure=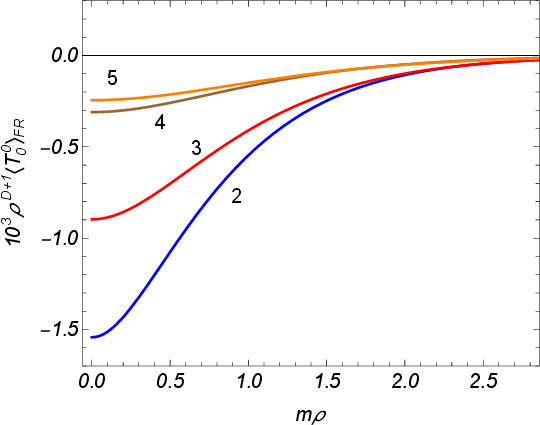,width=7.5cm,height=6.5cm}
\end{center}
\caption{The vacuum energy density in units of $1/\protect\rho ^{D+1}$ as a
function of $m\protect\rho $ for separate values of spatial dimension (the
numbers near the curves).}
\label{figT00}
\end{figure}

For the transformation of the energy density we use the representation
\begin{equation}
\left\vert K_{1/2-i\omega }(\lambda \rho )\right\vert ^{2}=2\int_{0}^{\infty
}du\,\cos \left( 2\omega u\right) K_{1}\left( 2\lambda \rho \cosh u\right) ,
\label{IntKK3}
\end{equation}%
which is a direct consequence of Eq. (\ref{RepKK3}) with $\rho ^{\prime
}=\rho $. Inserting in (\ref{Tmu2}) with $\mu =0$ and evaluating the
integral with the help of (\ref{IntK}) one finds%
\begin{equation}
\langle T_{0}^{0}\rangle _{\mathrm{FR}}=\frac{2Nm^{D+1}}{\left( 2\pi \right)
^{\frac{D+3}{2}}}\int_{0}^{\infty }du\,\frac{u^{2}-\pi ^{2}/4}{\left(
u^{2}+\pi ^{2}/4\right) ^{2}}f_{\frac{D+1}{2}}\left( 2m\rho \cosh u\right)
\cosh u.  \label{T00alt}
\end{equation}%
Yet another representation is obtained by using
\begin{equation}
\frac{u^{2}-\pi ^{2}/4}{\left( u^{2}+\pi ^{2}/4\right) ^{2}}=-\partial _{u}%
\frac{u}{u^{2}+\pi ^{2}/4},  \label{rel}
\end{equation}%
in (\ref{T00alt}) and integrating by parts. By taking into account the
relations%
\begin{eqnarray}
f_{\nu }^{\prime }(z) &=&-zf_{\nu +1}(z),  \notag \\
z^{2}f_{\nu +1}(z) &=&f_{\nu -1}(z)+2\nu f_{\nu }(z),  \label{fnurel}
\end{eqnarray}%
that gives%
\begin{equation}
\langle T_{0}^{0}\rangle _{\mathrm{FR}}=-\frac{2Nm^{D+1}}{\left( 2\pi
\right) ^{\frac{D+3}{2}}}\int_{0}^{\infty }du\,\frac{u\sinh u}{u^{2}+\pi
^{2}/4}\,\left[ f_{\frac{D-1}{2}}(2m\rho \cosh u)+Df_{\frac{D+1}{2}}(2m\rho
\cosh u)\right] .  \label{T00alt2}
\end{equation}%
Again, this shows that the energy density is negative. Figure \ref{figT00}
presents the dependence of the dimensionless product $\rho ^{D+1}\langle
T_{0}^{0}\rangle _{\mathrm{FR}}$ on $m\rho $ in spatial dimensions $D=2,3,4,5
$.

\begin{figure}[tbph]
\begin{center}
\epsfig{figure=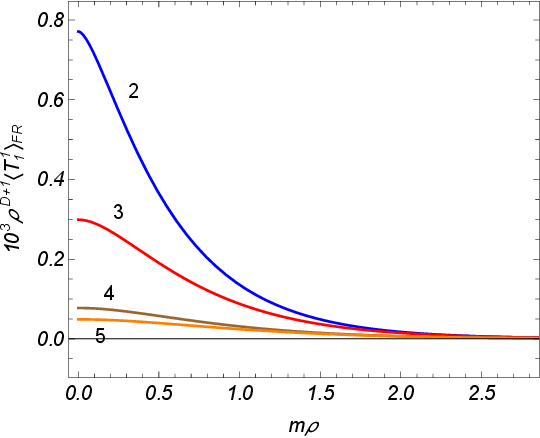,width=7.5cm,height=6.5cm}
\end{center}
\caption{The vacuum stress $\langle T_{1}^{1}\rangle _{\mathrm{FR}}$ in
units of $1/\protect\rho ^{D+1}$ as a function of $m\protect\rho $ in
different spatial dimensions.}
\label{figT11}
\end{figure}

For the vacuum stresses we can proceed in a similar way. For the stress $%
\langle T_{1}^{1}\rangle _{\mathrm{FR}}$ we use the representation%
\begin{equation}
\mathrm{Im}\left[ K_{\frac{1}{2}+i\omega }\left( \lambda \rho \right) K_{%
\frac{1}{2}-i\omega }^{\prime }\left( \lambda \rho \right) \right] =\frac{1}{%
\lambda \rho }\int_{0}^{\infty }dt\,\sin \left( 2\omega t\right) K_{1}\left(
2\lambda \rho \cosh t\right) \tanh t,  \label{IntKK4}
\end{equation}%
obtained from (\ref{RepKK4}). The following steps are based on (\ref{IntK})
and are similar to those for the FC and energy density. The final result
reads%
\begin{equation}
\langle T_{1}^{1}\rangle _{\mathrm{FR}}=\frac{2Nm^{D+1}}{\left( 2\pi \right)
^{\frac{D+3}{2}}}\int_{0}^{\infty }du\,\frac{u\sinh u}{u^{2}+\pi ^{2}/4}f_{%
\frac{D+1}{2}}\left( 2m\rho \cosh u\right) .  \label{T11alt}
\end{equation}%
For the stresses $\langle T_{l}^{l}\rangle _{\mathrm{FR}}$, $l=2,\ldots ,D$,
we employ the representation (\ref{IntKK2}) and the subsequent evaluation
shows that (no summation over $l$)%
\begin{equation}
\langle T_{l}^{l}\rangle _{\mathrm{FR}}=\langle T_{1}^{1}\rangle _{\mathrm{FR%
}},\;l=2,\ldots ,D.  \label{Tll}
\end{equation}%
Hence, we conclude that the vacuum stresses are isotropic for massive fields
as well. From (\ref{T11alt}) it follows that they are positive and, hence,
the corresponding vacuum pressure is negative. By using the representations
obtained, we can prove the trace relation $\langle T_{\mu }^{\mu }\rangle _{%
\mathrm{FR}}=m\left\langle \bar{\psi}\psi \right\rangle _{\mathrm{FR}}$.
This serves as an additional check for the calculations provided above. In
Figure \ref{figT11} we display the function $\rho ^{D+1}\langle
T_{1}^{1}\rangle _{\mathrm{FR}}$ versus $m\rho $. The numbers near the
curves are the values of the spatial dimension $D$.

The representations given in this section are further simplified for a
massless field. By taking into account that $f_{\nu }(z)\sim 2^{\nu
-1}\Gamma \left( \nu \right) z^{-2\nu }$ for $z\rightarrow 0$, we get%
\begin{equation}
\langle T_{\mu }^{\nu }\rangle _{\mathrm{FR}}=\frac{2N\Gamma \left( \frac{D+1%
}{2}\right) }{\left( 4\pi \right) ^{\frac{D+3}{2}}\rho ^{D+1}}%
\int_{0}^{\infty }du\,\frac{u\sinh u}{\left( u^{2}+\pi ^{2}/4\right) \cosh
^{D+1}u}\,\mathrm{diag}\left( -D,1,1,\ldots ,1\right) ,  \label{Tmum0c}
\end{equation}%
and FC vanishes. The integral in (\ref{Tmum0c}) is strongly convergent and
this representation is well adapted for numerical evaluation. For large
values of $m\rho $, by using $f_{\nu }(x)\approx \sqrt{\pi /2}x^{-\nu
-1/2}e^{-x}$ for $x\gg 1$, the dominant contribution to the integrals in (%
\ref{T00alt2}) and (\ref{T11alt}) comes from the region near the lower limit
and it can be seen that, to the leading order, the asymptotics (\ref{T00as1}%
) are obtained.

\section{VEVs in backgrounds conformally related to Rindler spacetime}

\label{sec:ConfRel}

Having the VEVs for a massless Dirac field in the Fulling-Rindler vacuum we
can generate the corresponding VEVs in the problems where the background
geometry is conformally related to the Rindler spacetime. Some examples for
a conformally coupled massless scalar field in 4-dimensional spacetime are
summarized in \cite{Birr82,Cand79}. In Ref. \cite{Saha04c} the brane induced
vacuum energy-momentum tensor is investigated in dS spacetime by using the
conformal relation with the problem of a planar boundary uniformly
accelerated through the Fulling-Rindler vacuum and the results from \cite%
{Saha02}.

Let $\bar{g}_{\mu \nu }$ be the metric tensor for a spacetime conformally
related to the Rindler metric, $d\bar{s}^{2}=\bar{g}_{\mu \nu }dx^{\mu
}dx^{\nu }=\Omega ^{2}(x)ds_{\mathrm{R}}^{2}$. The following relation takes
place for the respective mean energy-momentum tensors \cite{Birr82}:%
\begin{equation}
\langle \bar{T}_{\mu }^{\nu }\rangle _{\mathrm{CR}}=\sqrt{g/\bar{g}}\langle
T_{\mu }^{\nu }\rangle _{\mathrm{FR}}+\langle \bar{T}_{\mu }^{\nu }\rangle _{%
\mathrm{G}},  \label{Trel}
\end{equation}%
where $g$ and $\bar{g}$ are the determinants of the corresponding metric
tensors and $\langle T_{\mu }^{\nu }\rangle _{\mathrm{G}}$ is the
geometrical part that is completely determined by the geometrical
characteristics of the metric tensor $\bar{g}_{\mu \nu }$. The expectation
value in the left-hand side of Eq. (\ref{Trel}), denoted as $\langle \cdots
\rangle _{\mathrm{CR}}$, is taken for the state of quantum field conformally
related to the Fulling-Rindler vacuum in flat spacetime. A similar relation
is valid for the state of the field conformal to the Minkowski vacuum. By
taking into account that for the latter state the renormalized VEV of the
energy-momentum tensor vanishes, we see that $\langle \bar{T}_{\mu }^{\nu
}\rangle _{\mathrm{CM}}=$ $\langle \bar{T}_{\mu }^{\nu }\rangle _{\mathrm{G}%
} $, where CM stands for the state in the geometry with the metric $\bar{g}%
_{\mu \nu }$ that is conformal to the Minkowski vacuum in flat spacetime.
Hence, the relation (\ref{Trel}) can be written as%
\begin{equation}
\langle \bar{T}_{\mu }^{\nu }\rangle _{\mathrm{CR}}-\langle \bar{T}_{\mu
}^{\nu }\rangle _{\mathrm{CM}}=\sqrt{g/\bar{g}}\langle T_{\mu }^{\nu
}\rangle _{\mathrm{FR}}.  \label{Trel2}
\end{equation}%
The trace anomaly comes from the part $\langle \bar{T}_{\mu }^{\nu }\rangle
_{\mathrm{CM}}$ of the energy-momentum tensor. In odd-dimensional spacetimes
there is no trace anomaly and $\langle \bar{T}_{\mu }^{\nu }\rangle _{%
\mathrm{CM}}=0$.

As the first example we consider the conformal relation between the Rindler
spacetime and static spacetime with constant negative curvature spatial
sections. In order to see the relation between the metric tensors we
introduce angular coordinates $(\theta _{1},\theta _{2},\ldots ,\theta
_{D-2},\phi )$ in accordance with (see also \cite{Saha22}) $x^{l}=\rho
w^{l}\sinh r$, $l=2,\ldots ,D$, where $w^{i}=\cos \theta
_{i}\prod_{n=1}^{i-1}\sin \theta _{n}$ for $i=2,\ldots ,D-2$, $w^{D-1}=\cos
\phi \prod_{n=1}^{D-2}\sin \theta _{n}$, and $w^{D}=\sin \phi
\prod_{n=1}^{D-2}\sin \theta _{n}$, with $0\leq \theta _{l}\leq \pi $, $%
l=1,\ldots ,D-2$, $0\leq \phi \leq 2\pi $. The new time and radial
coordinates, $\eta $ and $r$, are defined by the relations
\begin{equation}
\tau =\frac{\eta }{\alpha },\;\rho =\frac{\alpha }{\cosh \left( r/\alpha
\right) -\sinh \left( r/\alpha \right) \cos \theta _{1}},  \label{Trans2}
\end{equation}%
and the parameter $\alpha $ determines the curvature radius of spatial
sections. In terms of the new coordinates, the Rindler line element is
presented as $ds_{\mathrm{R}}^{2}=\left( \rho /\alpha \right) ^{2}ds_{%
\mathrm{NC}}^{2}$, where%
\begin{equation}
ds_{\mathrm{NC}}^{2}=d\eta ^{2}-dr^{2}-\alpha ^{2}\sinh ^{2}\left( r/\alpha
\right) d\Omega _{D-1}^{2},  \label{ds2NC}
\end{equation}%
is the line element for a static spacetime with negative constant curvature
spatial foliation (static open universe). Here, $d\Omega _{D-1}^{2}$ is the
line element on a $(D-1)$-dimensional unit sphere. Hence, for the example
under consideration $\Omega (x)=\alpha /\rho $ and $\sqrt{g/\bar{g}}=\left(
\rho /\alpha \right) ^{D+1}$. Now, from (\ref{Trel}) we conclude that the
mean energy-momentum tensor for a massless Dirac field in the spacetime
given by the line element (\ref{ds2NC}) is given by%
\begin{equation}
\langle \bar{T}_{\mu }^{\nu }\rangle _{\mathrm{NC}}=\langle \bar{T}_{\mu
}^{\nu }\rangle _{\mathrm{G}}+\frac{A_{D}^{\mathrm{(ferm)}}}{\alpha ^{D+1}}%
\mathrm{diag}\left( -1,\frac{1}{D},\ldots ,\frac{1}{D}\right) .  \label{TNC}
\end{equation}%
Geometry is homogeneous and it is natural that the VEV does not depend on
the spacetime point. In Ref. \cite{Bunc78} it has been shown that (see also
Ref. \cite{Birr82}) for a conformally coupled scalar field the renormalized
VEV $\langle \bar{T}_{\mu }^{\nu }\rangle _{\mathrm{NC}}$ is zero in spatial
dimension $D=3$. A similar feature was assumed for spin 1/2 and 1 fields.

The static spacetime with negative constant curvature spatial sections is
conformally related to the Milne universe. The latter is flat and the
corresponding line element reads $ds_{\mathrm{Milne}}^{2}=e^{2\eta /\alpha
}ds_{\mathrm{NC}}^{2}=\left( \alpha e^{\eta /\alpha }/\rho \right) ^{2}ds_{%
\mathrm{R}}^{2}$. For the synchronous time coordinate in the Milne universe
one has $t=\alpha e^{\eta /\alpha }$ and the line element is written as
\begin{equation}
ds_{\mathrm{Milne}}^{2}=dt^{2}-t^{2}\left[ \alpha ^{-2}dr^{2}+\sinh
^{2}\left( r/\alpha \right) d\Omega _{D-1}^{2}\right] .  \label{ds2Milne}
\end{equation}%
The counterpart of the Fulling-Rindler vacuum in the Milne universe is the
conformal vacuum. The latter is different from the adiabatic vacuum being
the counterpart of the inertial vacuum in the Minkowski spacetime. Hence,
for a massless Dirac field in the conformal vacuum of the Milne universe the
expectation value of the energy-momentum tensor reads%
\begin{equation}
\langle \bar{T}_{\mu }^{\nu }\rangle _{\mathrm{Milne}}=\frac{A_{D}^{\mathrm{%
(ferm)}}}{t^{D+1}}\mathrm{diag}\left( -1,\frac{1}{D},\ldots ,\frac{1}{D}%
\right) .  \label{TMilne}
\end{equation}%
The corresponding result for a conformally coupled massless scalar field is
presented in \cite{Saha22}.

The third example of the conformal relation we are going to study is the dS
spacetime foliated by negative curvature spatial sections. The corresponding
line element reads%
\begin{equation}
ds_{\mathrm{dS}}^{2}=dt^{2}-\sinh ^{2}(t/\alpha )\left[ dr^{2}+\alpha
^{2}\sinh ^{2}\left( r/\alpha \right) d\Omega _{D-1}^{2}\right] .
\label{ds2dS}
\end{equation}%
Introducing the conformal time $\eta $ as $e^{\eta /\alpha }=\tanh
(t/2\alpha )$, the conformal relations%
\begin{equation}
ds_{\mathrm{dS}}^{2}=\frac{ds_{\mathrm{NC}}^{2}}{\sinh ^{2}\left( \eta
/\alpha \right) }=\left[ \frac{\alpha }{\rho \sinh \left( \eta /\alpha
\right) }\right] ^{2}ds_{\mathrm{R}}^{2},  \label{ds2c}
\end{equation}%
are seen. For this example $\sqrt{g/\bar{g}}=[\rho \sinh \left( |\eta
|/\alpha \right) /\alpha ]^{D+1}$. In dS spacetime the conformal counterpart
of the Minkowski vacuum is the Bunch-Davies vacuum. The latter is maximally
symmetric and the corresponding VEV has the structure $\langle \bar{T}_{\mu
}^{\nu }\rangle _{\mathrm{CM}}=\langle \bar{T}_{\mu }^{\nu }\rangle _{%
\mathrm{BD}}=\mathrm{const}\cdot \delta _{\mu }^{\nu }$ (the conformal
relation between the dS and Rindler spacetimes has been recently used in
\cite{Higu18} to clarify the entanglement structure in the Bunch-Davies
vacuum state). For even values of $D$ the trace anomaly is absent and $%
\langle \bar{T}_{\mu }^{\nu }\rangle _{\mathrm{BD}}=0$. For dS spacetime,
the conformal counterpart of the FR vacuum is the hyperbolic vacuum.
Denoting the respective VEV by $\langle \bar{T}_{\mu }^{\nu }\rangle _{%
\mathrm{H}}$, in terms of the synchronous time coordinate, from (\ref{Trel2}%
) we get%
\begin{equation}
\langle \bar{T}_{\mu }^{\nu }\rangle _{\mathrm{H}}-\langle \bar{T}_{\mu
}^{\nu }\rangle _{\mathrm{BD}}=\frac{A_{D}^{\mathrm{(ferm)}}}{\left[ \alpha
\sinh \left( t/\alpha \right) \right] ^{D+1}}\mathrm{diag}\left( -1,\frac{1}{%
D},\ldots ,\frac{1}{D}\right) .  \label{TdS}
\end{equation}%
The corresponding relation for a massless scalar field has been recently
discussed in \cite{Saha22} (for boundary induced effects in the hyperbolic
vacuum of dS spacetime for a massive scalar field with general curvature
coupling see \cite{Saha21}).

We can also find the expectation value of the energy-momentum tensor for a
massless Dirac field in the static vacuum of dS spacetime. The latter is the
vacuum state for an observer with fixed radial coordinate $r_{s}$ and
angular coordinates $(\theta _{1},\theta _{2},\ldots ,\theta _{D-2},\phi )$.
The respective line element has the form%
\begin{equation}
ds_{\mathrm{dS}}^{2}=\left( 1-r_{s}^{2}/\alpha ^{2}\right) d\eta ^{2}-\frac{%
dr_{s}^{2}}{1-r_{s}^{2}/\alpha ^{2}}-r_{s}^{2}d\Omega _{D-1}^{2}.
\label{ds2d}
\end{equation}%
Introducing a new radial coordinate $r$ in accordance with $r_{s}/\alpha
=\tanh \left( r/\alpha \right) $, the following conformal relations are
obtained:%
\begin{equation}
ds_{\mathrm{dS}}^{2}=\frac{ds_{\mathrm{NC}}^{2}}{\cosh ^{2}\left( r/\alpha
\right) }=\frac{\alpha ^{2}-r_{s}^{2}}{\rho ^{2}}ds_{\mathrm{R}}^{2},
\label{ds2e}
\end{equation}%
where, as it is seen from (\ref{Trans2}), we have%
\begin{equation}
\;\rho =\frac{\alpha \sqrt{1-r_{s}^{2}/\alpha ^{2}}}{1-\frac{r_{s}}{\alpha }%
\cos \theta _{1}}.  \label{rors}
\end{equation}%
Now, from (\ref{Trel2}) one obtains%
\begin{equation}
\langle \bar{T}_{\mu }^{\nu }\rangle _{\mathrm{St}}-\langle \bar{T}_{\mu
}^{\nu }\rangle _{\mathrm{BD}}=\frac{A_{D}^{\mathrm{(ferm)}}}{\left( \alpha
^{2}-r_{s}^{2}\right) ^{\frac{D+1}{2}}}\mathrm{diag}\left( -1,\frac{1}{D}%
,\ldots ,\frac{1}{D}\right) ,  \label{Tst}
\end{equation}%
where $\langle \bar{T}_{\mu }^{\nu }\rangle _{\mathrm{St}}$ is the VEV for
the static vacuum in dS spacetime. For $D=3$ one has $A_{D}^{\mathrm{(ferm)}%
}=17/(1920\pi ^{2})$ and (\ref{Tst}) is reduced to the result given in \cite%
{Cand79}. Note that in 4-dimensional dS spacetine and for the Bunch-Davies
vacuum $\langle \bar{T}_{\mu }^{\nu }\rangle _{\mathrm{BD}}=11\delta _{\mu
}^{\nu }/\left( 1920\pi ^{2}\alpha ^{4}\right) $.

\section{Conclusion}

\label{sec:Conc}

We have studied the local properties of the Fulling-Rindler vacuum for a
massive Dirac field in general number of spatial dimensions. As
characteristics of the properties the FC and the expectation value of the
energy-momentum tensor are considered. For evaluating the corresponding VEVs
the mode summation technique is employed with combination of the
point-splitting regularization procedure. As the first step we have
generalized the result of Ref. \cite{Cand78} for the complete set of
fermionic normal modes in Rindler spacetime for a massive field in $(D+1)$%
-dimensional spacetime. The positive and negative energy mode functions with
respect to the Rindler Killing vector $\partial _{\tau }$ are given by
expressions (\ref{psip2}) and (\ref{psim}) with the normalization
coefficient (\ref{Nsig}). The constant spinors $\chi _{\eta }^{(j)}(\mathbf{k%
})$, $j=+,-$, obey the relations (\ref{Relxi1})-(\ref{xirel}), with $(+)$ in
superscript replace by $(j)$, and no specific form of the Dirac matrices has
been used in deriving the modes. Both the FC and the VEV of the
energy-momentum tensor are presented as coincidence limits of two-point
functions and those limits are divergent. The Rindler spacetime is flat and
the renormalization is reduced to the subtraction of the corresponding
quantities for the Minkowski vacuum, transformed to the Rindler coordinate
system. In Appendix \ref{sec:AppMink} we have presented the Minkowskian VEVs
in the form adapted for the subtraction.

The renormalized FC is given by the expression (\ref{FCren1}). An
alternative representation is provided by Eq. (\ref{FCalt}). The FC vanishes
for massless fields and is negative for massive fields. The dependence of
the combination $m^{-D}\left\langle \bar{\psi}\psi \right\rangle _{\mathrm{FR%
}}$ on the mass of the field and on the acceleration $1/\rho $ appears in
the form of a single argument $m\rho $. For large accelerations
corresponding to small values of that argument the FC behaves as $m/\rho
^{D-1}$. This limit corresponds to spacetime points near the Rindler
horizon. In the opposite limit of small accelerations the FC is suppressed
by the factor $\left( m\rho \right) ^{-\frac{D+3}{2}}e^{-2m\rho }$.

Another important local characteristic of the vacuum state is the
expectation value of the energy-momentum tensor. It is diagonal for the
Fulling-Rindler vacuum and the respective components are given by (\ref{Tmu2}%
) with the functions (\ref{Fmuc}) for separate components. Simpler formulas
are obtained by using the integral representations of the products for the
modified Bessel functions, given in Appendix \ref{sec:AppMink}. The vacuum
energy density is expressed as (\ref{T00alt}) (or equivalently by (\ref%
{T00alt2})) and is negative. The vacuum stresses are isotropic and are given
by (\ref{T11alt}). The respective vacuum pressures are negative. As
expected, the VEV of the energy-momentum tensor obeys the covariant
conservation equation and the trace relation with the FC. For a massless
field the general expression is simplified to the form (\ref{Tmum0b}) with
the coefficient $A_{D}^{\mathrm{(ferm)}}$ defined by (\ref{ADf}). The
massless Dirac field is conformally invariant in all spatial dimensions and
the respective energy-momentum tensor is traceless. The background geometry
is flat and the trace anomaly is absent. Near the Rindler horizon,
corresponding to large accelerations, the effect of the mass on the vacuum
energy-momentum tensor is weak and the leading term in the asymptotic
expansion coincides withe the VEV for a massless field with behavior $1/\rho
^{D+1}$. For small accelerations the decay of the vacuum energy-momentum
tensor for massive fields is exponential. The suppression factor is the same
as that for the FC. It should be noted that the isotropy of the stresses for
the Fulling-Rindler vacuum does not follow from the symmetry of the problem.
In the case of a scalar field, in general, the stress $\langle
T_{1}^{1}\rangle _{\mathrm{FR}}$ differs from those along the other
directions, $\langle T_{l}^{l}\rangle _{\mathrm{FR}}$, $l=2,\ldots ,D$. The
equality takes place only for a conformally coupled massless field.

One can interpret the tensor $-\langle T_{\mu }^{\nu }\rangle _{\mathrm{FR}}$
as the energy-momentum tensor for the Minkowski vacuum seen by a uniformly
accelerating observer. For a massless fermionic field the form of that
tensor follows from (\ref{Tmum0b}) and it describes a radiation type source
with the equation of state $P=\varepsilon /D$, where (no summation over $l$)
$P=\langle T_{l}^{l}\rangle _{\mathrm{FR}}$, $l=1,\ldots ,D$, is the
pressure and $\varepsilon =-\langle T_{0}^{0}\rangle _{\mathrm{FR}}$ is the
energy density. Both these quantities are positive. By taking into account
that the energy of the fermionic mode with a given $\omega $, measured by an
observer with proper acceleration $1/\rho $, is given by $\varepsilon _{\rho
}=\omega /\rho $, we can interpret the factor $[e^{2\pi \rho \varepsilon
_{\rho }}-\left( -1\right) ^{D}]^{-1}$ in the integrand of (\ref{ADf}) as an
indication for the thermal nature of the energy distribution with the Unruh
temperature $T_{U}=1/(2\pi \rho )$. An interesting point to be mentioned is
that the thermal factor is of the Fermi-Dirac type only in odd number of
spatial dimensions $D$. In even dimensional spaces the Bose-Einstein type
factor appears. For a conformally coupled massless scalar field one has an
inverted situation: the thermal factor is of the Bose-Einstein type for odd $%
D$ and of the Fermi-Dirac type in the case of even $D$. A similar feature
was observed in the response of uniformly accelerating particle detectors.

In accordance of the equivalence principle, the local effects of the
gravitational filed on the properties of a physical system can be modeled by
considering the system in non-inertial reference frame with an appropriate
acceleration. In this context, the results presented above shed light on the
influence of the gravitational field on the local properties of the fermionic
vacuum in different numbers of spatial dimensions. The Rindler spacetime
also approximates the near-horizon geometry of a large class of black holes
and the VEVs described above give the leading terms in the respective
expansions over the distance from the horizon. Other applications are based
on the conformal relations of the Rindler spacetime with curved geometries
in gravitational physics and cosmology. For a massless field the fermionic
expectation values in those geometries are obtained by using the conformal
relation between the VEVs. Some examples are presented in the discussion
above.

\section*{Acknowledgments}

V.Kh.K. was supported by the grants No. 22AA-1C002 and No. 21AG-1C069 of the
Science Committee of the Ministry of Education, Science, Culture and Sport
RA. A.A.S. was supported by the grant No. 21AG-1C047 of the Science
Committee of the Ministry of Education, Science, Culture and Sport RA.
V.Kh.K. and A.A.S. were partially supported by the ANSEF grant
23AN:PS-hepth-2889. A.A.S. gratefully acknowledges the hospitality of the
INFN, Laboratori Nazionali di Frascati (Frascati, Italy), where a part of
this work was done.

\appendix

\section{Integral representations for the products of Macdonald functions}

\label{sec:App1}

In this appendix we present integral representations for the products of the
Macdonald functions appearing in the expressions for the fermionic VEVs. The
consideration will be based on the integral representation \cite{Erde63}
\begin{equation}
K_{\mu }(Z)K_{\nu }(z)=\int_{-\infty }^{+\infty }du\,e^{-(\mu -\nu )u}\left(
\frac{Ze^{u}+ze^{-u}}{Ze^{-u}+ze^{u}}\right) ^{(\mu +\nu )/2}K_{\mu +\nu
}\left( \sqrt{Z^{2}+z^{2}+2Zz\cosh \left( 2u\right) }\right) ,  \label{RepKK}
\end{equation}%
for $\mathrm{Re}\,Z>0$, $\mathrm{Re}\,z>0$. Taking $\mu =-\nu =1/2-i\omega $
we get%
\begin{equation}
K_{\frac{1}{2}-i\omega }(\lambda \rho )K_{\frac{1}{2}-i\omega }(\lambda \rho
^{\prime })=\frac{1}{2}\int_{-\infty }^{+\infty }du\,e^{i\omega
u-u/2}K_{0}\left( \lambda \Delta \left( \rho ,\rho ^{\prime },u\right)
\right) ,  \label{RepKK1}
\end{equation}%
where%
\begin{equation}
\Delta \left( \rho ,\rho ^{\prime },u\right) =\sqrt{\rho ^{2}+\rho ^{\prime
2}+2\rho \rho ^{\prime }\cosh u}.  \label{Del}
\end{equation}

Next, we take $\mu =1/2-i\omega $, $\nu =1/2+i\omega $ in (\ref{RepKK}):%
\begin{equation}
K_{\frac{1}{2}-i\omega }(\lambda \rho )K_{\frac{1}{2}+i\omega }(\lambda \rho
^{\prime })=\frac{1}{2}\int_{-\infty }^{+\infty }du\,e^{i\omega u}\left(
\frac{\rho e^{u}+\rho ^{\prime }}{\rho ^{\prime }e^{u}+\rho }\right)
^{1/2}K_{1}\left( \lambda \Delta \left( \rho ,\rho ^{\prime },u\right)
\right) .  \label{RepKK3}
\end{equation}%
The expression for the $_{1}^{1}$-stress contains the product $K_{\frac{1}{2}%
+i\omega }\left( \lambda \rho \right) K_{\frac{1}{2}-i\omega }^{\prime
}\left( \lambda \rho ^{\prime }\right) $. In order to find the integral
representation we first take the derivative of (\ref{RepKK}) with respect to
$\rho ^{\prime }$ and then $\mu =1/2+i\omega $, $\nu =1/2-i\omega $. That
gives%
\begin{eqnarray}
K_{\frac{1}{2}+i\omega }(\lambda \rho )K_{\frac{1}{2}-i\omega }^{\prime
}(\lambda \rho ^{\prime }) &=&\frac{1}{2}\int_{-\infty }^{+\infty }du\,\frac{%
e^{u/2-i\omega u}}{\rho +\rho ^{\prime }e^{u}}\left[ -\rho \sinh u\frac{%
K_{1}\left( \lambda \Delta \left( \rho ,\rho ^{\prime },u\right) \right) }{%
\lambda \Delta \left( \rho ,\rho ^{\prime },u\right) }\right.  \notag \\
&&\left. +K_{1}^{\prime }\left( \lambda \Delta \left( \rho ,\rho ^{\prime
},u\right) \right) \left( \rho ^{\prime }+\rho \cosh u\right) \right] .
\label{RepKK4}
\end{eqnarray}

From (\ref{RepKK1}) it follows that%
\begin{equation}
K_{0}\left( \lambda \Delta \left( \rho ,\rho ^{\prime },u\right) \right) =%
\frac{e^{\frac{u}{2}}}{\pi }\int_{-\infty }^{+\infty }d\omega \,e^{-i\omega
u}K_{\frac{1}{2}-i\omega }(\lambda \rho )K_{\frac{1}{2}-i\omega }(\lambda
\rho ^{\prime }).  \label{RepK0}
\end{equation}%
This representation will be used for the transformation of the FC in
Minkowski vacuum. The corresponding expression contains the function $%
K_{0}\left( \lambda \sqrt{\rho ^{2}+\rho ^{\prime 2}-2\rho \rho ^{\prime
}\cosh \Delta \tau }\right) $. Note that we can write%
\begin{equation*}
\rho ^{2}+\rho ^{\prime 2}-2\rho \rho ^{\prime }\cosh \Delta \tau =\Delta
\left( \rho ,\rho ^{\prime },\Delta \tau \pm i\pi \right) ,
\end{equation*}%
and from (\ref{RepK0}) one gets%
\begin{equation}
K_{0}\left( \lambda \sqrt{\rho ^{2}+\rho ^{\prime 2}-2\rho \rho ^{\prime
}\cosh \Delta \tau }\right) =\frac{\pm i}{\pi }e^{\frac{\Delta \tau }{2}%
}\int_{-\infty }^{+\infty }d\omega \,e^{-i\omega \Delta \tau \pm \pi \omega
}K_{\frac{1}{2}-i\omega }(\lambda \rho )K_{\frac{1}{2}-i\omega }(\lambda
\rho ^{\prime }).  \label{RepK01}
\end{equation}%
Summing the representations with upper and lower signs we obtain%
\begin{equation}
K_{0}\left( \lambda \sqrt{\rho ^{2}+\rho ^{\prime 2}-2\rho \rho ^{\prime
}\cosh \Delta \tau }\right) =\frac{2}{\pi }e^{\frac{\Delta \tau }{2}%
}\int_{0}^{\infty }d\omega \,\sinh \left( \pi \omega \right) \,\mathrm{Im}%
\left[ e^{i\omega \Delta \tau }K_{\frac{1}{2}+i\omega }(\lambda \rho )K_{%
\frac{1}{2}+i\omega }(\lambda \rho ^{\prime })\right] .  \label{RepK02}
\end{equation}%
Here, the imaginary part in the integrand is understood as
\begin{equation}
\mathrm{Im}\left[ e^{i\omega \Delta \tau }f(i\omega )\right] =\frac{%
e^{i\omega \left( \Delta \tau +i\varepsilon \right) }f(i\omega )-e^{i\omega
\left( \Delta \tau -i\varepsilon \right) }f(i\omega )}{2i}.  \label{Imrel}
\end{equation}%
with small $\varepsilon >0$.

\section{Hadamard function and FC for the Minkowski vacuum}

\label{sec:AppMink}

In order to obtain a finite result for the FC (\ref{FC2}) we subtract the
corresponding quantity for the Minkowski vacuum. The background geometry is
flat and the divergences are the same as those in the Minkowski vacuum and
this subtraction is sufficient. Here we obtain a representation for the
Minkowskian FC adapted for the subtraction procedure. The Minkowskian line
element is presented as $ds_{\mathrm{M}}^{2}=dt^{2}-d\mathbf{z}^{2}$, where $%
\mathbf{z}=(x^{1},\mathbf{x})$. The corresponding Dirac equation reads $%
\left( i\gamma ^{(\mu )}\partial _{\mu }-m\right) \psi _{\mathrm{M}}(x)=0$.
Similar to the case of the Rindler geometry, we present the field in the
form
\begin{equation}
\psi _{\mathrm{M}}(x)=\left( i\gamma ^{(\nu )}\partial _{\nu }+m\right)
\varphi _{\mathrm{M}}(x),  \label{psiM}
\end{equation}%
where the new function obeys the equation $\left( \eta ^{\mu \nu }\partial
_{\mu }\partial _{\nu }+m^{2}\right) \varphi _{\mathrm{M}}(x)=0$ with the
Minkowski metric tensor $\eta ^{\mu \nu }=\mathrm{diag}(1,-1,\ldots ,-1)$.
The solution for the latter equation is expressed as $\varphi _{\mathrm{M}%
}(x)=u_{\mathrm{M}}(\mathbf{K})e^{-i\Omega t+i\mathbf{K}\cdot \mathbf{z}}$,
where $\mathbf{K}=(K^{1},\mathbf{k})$, $\mathbf{K}\cdot \mathbf{z}%
=K^{1}x^{1}+\mathbf{kx}$, and $\Omega =\sqrt{K^{2}+m^{2}}$. Now, from the
relation (\ref{psiM}) for the normalized positive and negative energy
fermionic modes, realizing the Minkowski vacuum, we get
\begin{equation}
\psi _{\mathrm{M}\sigma }^{(\pm )}(x)=\frac{\chi _{\mathrm{M}\eta }^{(\pm )}(%
\mathbf{K})}{\left( 2\pi \right) ^{D/2}}e^{\mp i\Omega t+i\mathbf{K}\cdot
\mathbf{z}},  \label{psiM2}
\end{equation}%
where the constant spinors $\chi _{\mathrm{M}\eta }^{(\pm )}(\mathbf{K})$
are the eigenspinors of the projection operator
\begin{equation}
P_{\mathrm{M}}^{(\pm )}(\mathbf{K})=\frac{1}{2}\left( 1\pm \gamma ^{(0)}%
\frac{\boldsymbol{\gamma \,}\mathbf{K}+m}{\Omega }\right) ,  \label{PM}
\end{equation}%
with the properties $P_{\mathrm{M}}^{(\pm )2}(\mathbf{K})=P_{\mathrm{M}%
}^{(\pm )}(\mathbf{K})$, $P_{\mathrm{M}}^{(\pm )\dagger }(\mathbf{K})=P_{%
\mathrm{M}}^{(\pm )}(\mathbf{K})$, and $\boldsymbol{\gamma \,}\mathbf{K}%
=\sum_{l=1}^{D}\gamma ^{(l)}K^{l}$. The corresponding relation reads $P_{%
\mathrm{M}}^{(\pm )}(\mathbf{K})\chi _{\mathrm{M}\eta }^{(\pm )}(\mathbf{K}%
)=\chi _{\mathrm{M}\eta }^{(\pm )}(\mathbf{K})$. The respective
orthogonality and completeness relations have the form (compare with (\ref%
{Relxi2}), (\ref{xirel}))%
\begin{eqnarray}
\chi _{\mathrm{M}\eta }^{(\pm )\dagger }(\mathbf{K})\chi _{\mathrm{M}\eta
^{\prime }}^{(\pm )}(\mathbf{K}) &=&\delta _{\eta \eta ^{\prime }},  \notag
\\
\sum_{\eta }\chi _{\mathrm{M}\eta \alpha }^{(\pm )}(\mathbf{K})\chi _{%
\mathrm{M}\eta \beta }^{(\pm )\dagger }(\mathbf{K}) &=&P_{\mathrm{M}\alpha
\beta }^{(\pm )}(\mathbf{K}).  \label{xiMrel}
\end{eqnarray}

First we consider the positive frequency Wightman function $S_{\mathrm{M}%
}^{+}(x,x^{\prime })=\left\langle \psi (x)\bar{\psi}(x^{\prime
})\right\rangle _{\mathrm{M}}$. Substituting the mode functions (\ref{psiM2}%
) in the corresponding mode sum formula and using the relation (\ref{xiMrel}%
) for the summation over $\eta $, we get%
\begin{equation}
S_{\mathrm{M}}^{+}(x,x^{\prime })=\frac{1}{2}\int \frac{d\mathbf{K}}{\left(
2\pi \right) ^{D}}\frac{e^{i\mathbf{K}\cdot \Delta \mathbf{z}}}{\Omega }%
\left( m+\gamma ^{(0)}\Omega -\boldsymbol{\gamma \,}\mathbf{K}\right)
e^{-i\Omega \Delta t}.  \label{SMp}
\end{equation}%
The respective expression for the negative frequency Wightman function $S_{%
\mathrm{M}}^{-}(x,x^{\prime })=\left\langle \bar{\psi}(x^{\prime })\psi
(x)\right\rangle _{\mathrm{M}}$ is obtained from (\ref{SMp}) by the
replacements $\Omega \rightarrow -\Omega $ and $\left( \cdots \right)
\rightarrow \left( \cdots \right) ^{T}$, where $T$ stands for the
transponation. The formula (\ref{SMp}) gives the Wightman function for the
inertial vacuum in the Minkowskian coordinates. Though the FC $\left\langle
\bar{\psi}\psi \right\rangle _{\mathrm{M}}$ is a scalar with respect to the
transformation to the Rindler coordinates, that is not the case for
two-point functions and also for the corresponding traces. The Wightman
function transforms as a product of two spinors $\psi (x)$ and $\bar{\psi}%
(x^{\prime })$. The coordinate transformation $(t,x^{1},\mathbf{x}%
)\rightarrow (\tau ,\rho ,\mathbf{x})$, given by (\ref{trans}), induces a
local Lorentz transformation $L(x)$ (for a general discussion in curved
spacetime with an arbitrary number of spatial dimensions see, e.g., \cite%
{Park09}). Under the latter transformation the spinor transforms as $\psi _{%
\mathrm{MR}}=U(L(x))\psi _{\mathrm{M}}$, where the transformation matrix
reads
\begin{equation}
U(x)=e^{-\tau \gamma ^{(0)}\gamma ^{(1)}/2}=\cosh \left( \tau /2\right)
-\gamma ^{(0)}\gamma ^{(1)}\sinh \left( \tau /2\right) .  \label{Utr}
\end{equation}%
By taking into account that $\bar{\psi}_{\mathrm{MR}}=\bar{\psi}_{\mathrm{M}%
}U^{-1}(L(x))$, for the Wightman function in the Rindler coordinates one
obtains $S_{\mathrm{MR}}^{+}(x,x^{\prime })=U(L(x))S_{\mathrm{M}%
}^{+}(x,x^{\prime })U^{-1}(L(x^{\prime }))$. Combining this with (\ref{SMp}%
), for the trace of the Wightman functions we get
\begin{equation}
\mathrm{Tr}(S_{\mathrm{MR}}^{j}(x,x^{\prime }))=j\frac{m}{2}\cosh \left(
\frac{\Delta \tau }{2}\right) \int \frac{d\mathbf{K}}{\left( 2\pi \right)
^{D}}\frac{e^{-ji\Omega \Delta t+i\mathbf{K}\cdot \Delta \mathbf{z}}}{\Omega
}.  \label{TrSj}
\end{equation}%
For the trace of the Hadamard function $S_{\mathrm{MR}}^{(1)}(x,x^{\prime
})=\sum_{j=+,-}jS_{\mathrm{MR}}^{j}(x,x^{\prime })$, required in the
evaluation of the FC, this gives $\mathrm{Tr}(S_{\mathrm{MR}%
}^{(1)}(x,x^{\prime }))=\cosh \left( \frac{\Delta \tau }{2}\right) \mathrm{Tr%
}(S_{\mathrm{M}}^{(1)}(x,x^{\prime }))$, where for the trace in the
Minkowskian coordinates one has%
\begin{equation}
\mathrm{Tr}(S_{\mathrm{M}}^{(1)}(x,x^{\prime }))=m\int \frac{d\mathbf{K}}{%
\left( 2\pi \right) ^{D}}\frac{e^{i\mathbf{K}\cdot \Delta \mathbf{z}}}{%
\Omega }\cos \left( \Omega \Delta t\right) .  \label{Str}
\end{equation}

In order to find a representation of the trace (\ref{Str}) convenient for
the subtraction, first we integrate over $K^{1}$. This gives%
\begin{equation}
\mathrm{Tr}(S_{\mathrm{M}}^{(1)}(x,x^{\prime }))=2Nm\int \frac{d\mathbf{k}}{%
\left( 2\pi \right) ^{D}}\,e^{i\mathbf{k}\cdot \Delta \mathbf{x}}K_{0}\left(
\lambda \sqrt{\left( \Delta x^{1}\right) ^{2}-\left( \Delta t\right) ^{2}}%
\right) ,  \label{Str1}
\end{equation}%
where $\lambda $ is given by (\ref{lam}). As the next step, we express the
Minkowskian coordinates $(t,x^{1})$ in terms of the Rindler coordinates $%
(\tau ,\rho )$ for the replacement%
\begin{equation}
\left( \Delta x^{1}\right) ^{2}-\left( \Delta t\right) ^{2}=\rho ^{2}+\rho
^{\prime 2}-2\rho \rho ^{\prime }\cosh \Delta \tau ,  \label{coorRel}
\end{equation}%
and use the formula (\ref{RepK02}). As a result, the following
representation is obtained%
\begin{equation}
\mathrm{Tr}(S_{\mathrm{M}}^{(1)}(x,x^{\prime }))=\frac{Nme^{\frac{\Delta
\tau }{2}}}{2^{D-2}\pi ^{D+1}}\int d\mathbf{k}\,e^{i\mathbf{k}\cdot \Delta
\mathbf{x}}\int_{0}^{\infty }d\omega \,\sinh \left( \pi \omega \right) \,%
\mathrm{Im}\left[ e^{i\omega \Delta \tau }K_{\frac{1}{2}+i\omega }(\lambda
\rho )K_{\frac{1}{2}+i\omega }(\lambda \rho ^{\prime })\right] .
\label{Str2}
\end{equation}%
The corresponding trace in the Rindler coordinates is obtained by adding in
the right-hand side the factor $\cosh \left( \Delta \tau /2\right) $. Note
that the integral in (\ref{Str1}) is further evaluated with the result in
the Rindler coordinates%
\begin{equation}
\mathrm{Tr}(S_{\mathrm{MR}}^{(1)}(x,x^{\prime }))=\frac{2Nm^{D}}{\left( 2\pi
\right) ^{\frac{D+1}{2}}}\cosh \left( \Delta \tau /2\right) f_{\frac{D-1}{2}%
}\left( m\sigma (x,x^{\prime })\right) ,  \label{SMtr3}
\end{equation}%
where $\sigma (x,x^{\prime })=\sqrt{\left( \Delta x^{1}\right) ^{2}+|\Delta
\mathbf{x}|^{2}-\left( \Delta t\right) ^{2}}$ determines the geodesic
distance between the points $x$ and $x^{\prime }$ and the function $f_{\nu
}(z)$ is defined by (\ref{fnu}). The formal expression of the FC for the
Minkowski vacuum is obtained taking the limit $x^{\prime }\rightarrow x$ in (%
\ref{Str2}).

\section{Representation for the trace of the Hadamard function}

\label{sec:App3}

In this appendix we will provide representations for the Wightman and
Hadamard functions in terms of the respective functions for the Minkowski
vacuum. A similar procedure for the Feynman function of a scalar field has
been presented in \cite{Cand76}. The starting point for the transformation
of $\mathrm{Tr}\left( S^{(1)}(x,x^{\prime })\right) $ will be the formula (%
\ref{FC2}). First of all we note that the Hadamard function can be written
in terms of the positive and negative energy Wightman functions as $%
S^{(1)}(x,x^{\prime })=\sum_{j=+,-}jS^{j}(x,x^{\prime })$, where the
respective matrix elements are defined by%
\begin{equation}
S_{\alpha \beta }^{+}(x,x^{\prime })=\left\langle \psi _{\alpha }(x)\bar{\psi%
}_{\beta }(x^{\prime })\right\rangle ,\;S_{\alpha \beta }^{-}(x,x^{\prime
})=\left\langle \bar{\psi}_{\alpha }(x^{\prime })\psi _{\beta
}(x)\right\rangle .  \label{Spm}
\end{equation}%
The expressions for the traces of the latter functions follow from (\ref{FC2}%
):%
\begin{equation}
\mathrm{Tr}\left( S^{j}(x,x^{\prime })\right) =-\frac{jNm}{2^{D-1}\pi ^{D+1}}%
\int_{0}^{\infty }d\omega \,\cosh \left( \pi \omega \right) e^{-ji\omega
\Delta \tau }\int d\mathbf{k\,}e^{i\mathbf{k}\cdot \Delta \mathbf{x}}\,%
\mathrm{Im}\left[ K_{\frac{1}{2}-i\omega }\left( \lambda \rho \right) K_{%
\frac{1}{2}-i\omega }\left( \lambda \rho ^{\prime }\right) \right] .
\label{Spm2}
\end{equation}%
By using the integral representation (\ref{RepKK1}), the integral over the
momentum $\mathbf{k}$ is evaluated by the formula%
\begin{equation}
\int d\mathbf{k\,}e^{i\mathbf{k}\cdot \Delta \mathbf{x}}K_{0}\left( \lambda
\Delta \left( \rho ,\rho ^{\prime },u\right) \right) =\left( 2\pi \right) ^{%
\frac{D-1}{2}}m^{D-1}f_{\frac{D-1}{2}}\left( m\Delta _{D}(u)\right) ,
\label{IntK0}
\end{equation}%
with the notations (\ref{fnu}) and%
\begin{equation}
\Delta _{D}(u)=\sqrt{\rho ^{2}+\rho ^{\prime 2}+2\rho \rho ^{\prime }\cosh
u+|\Delta \mathbf{x}|^{2}}.  \label{DelD}
\end{equation}%
This leads to the result%
\begin{equation}
\mathrm{Tr}\left( S^{j}(x,x^{\prime })\right) =-j\frac{2Nm^{D}}{\left( 2\pi
\right) ^{\frac{D+3}{2}}}\int_{0}^{\infty }d\omega \,\cosh \left( \pi \omega
\right) e^{-ji\omega \Delta \tau }\int_{-\infty }^{+\infty }du\,\sin \left(
\omega u\right) e^{-u/2}f_{\frac{D-1}{2}}\left( m\Delta _{D}(u)\right) .
\label{Spm3}
\end{equation}%
In this form and for real $\Delta \tau $ the change of the order of
integrations is not allowed.

\begin{figure}[tbph]
\begin{center}
\epsfig{figure=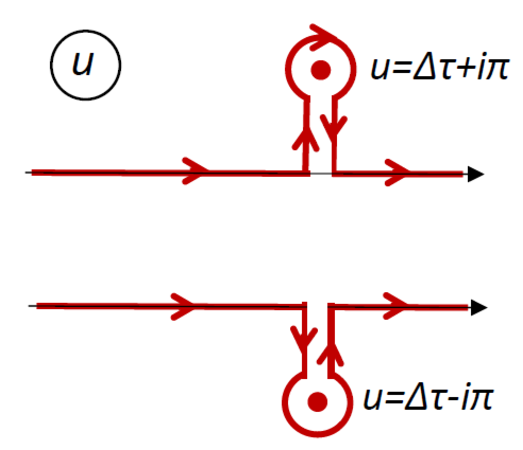,width=5.cm,height=4.cm}
\end{center}
\caption{The contours of integrations in (\protect\ref{Spm4}) for $j=+,-$.}
\label{Contfig}
\end{figure}

To make the change admissible, we will temporarily assume that $j\mathrm{Im\,%
}\Delta \tau <-\pi $. First integrating over $\omega $ one gets%
\begin{equation}
\int_{0}^{\infty }d\omega \,\cosh \left( \pi \omega \right) \sin \left(
\omega u\right) e^{-ji\omega \Delta \tau }=-\frac{1}{2}\sum_{\varkappa =\pm
1}\frac{\varkappa \left( \Delta \tau -\varkappa u\right) }{\left( \Delta
\tau -\varkappa u\right) ^{2}+\pi ^{2}}.  \label{Intom}
\end{equation}%
Plugging this in (\ref{Spm3}), passing to a new integration variable $%
u^{\prime }=\varkappa u$ and renoting again $u^{\prime }\rightarrow u$ we
can see that%
\begin{equation}
\mathrm{Tr}\left( S^{j}(x,x^{\prime })\right) =j\frac{2m^{D}N}{\left( 2\pi
\right) ^{\frac{D+3}{2}}}\int_{-\infty }^{+\infty }du\,\frac{\left( u-\Delta
\tau \right) \sinh \left( u/2\right) }{\left( u-\Delta \tau \right) ^{2}+\pi
^{2}}f_{\frac{D-1}{2}}\left( m\Delta _{D}(u)\right) .  \label{Spm4}
\end{equation}%
Recall that this representation is obtained for $\mathrm{Im\,}\Delta \tau
<-\pi $ for $j=+$ and $\mathrm{Im\,}\Delta \tau >\pi $ for $j=-$. An
analytical continuation is required in the limit $\mathrm{Im\,}\Delta \tau
\rightarrow 0$. The integrand has poles at $u=\Delta \tau \pm i\pi $ and
they are located below the integration contour (real axis) for $j=+$ and
above the contour for $j=-$. These locations dictate the deformation of the
integration contour in the limit $\mathrm{Im\,}\Delta \tau \rightarrow 0$.
The deformed contours are depicted in Figure \ref{Contfig} for $j=+$ (upper
contour) and for $j=-$ (lower contour). The integrals along the circles near
the points $u=\Delta \tau \pm \pi $ are expressed in terms of the respective
residues and we get%
\begin{eqnarray}
\mathrm{Tr}\left( S^{j}(x,x^{\prime })\right) &=&j\frac{2m^{D}N}{\left( 2\pi
\right) ^{\frac{D+3}{2}}}\left[ \pi \cosh \left( \Delta \tau /2\right) f_{%
\frac{D-1}{2}}\left( m\sigma (x,x^{\prime })\right) \right.  \notag \\
&&\left. +\int_{-\infty }^{+\infty }du\,\frac{\left( u-\Delta \tau \right)
\sinh \left( u/2\right) }{\left( u-\Delta \tau \right) ^{2}+\pi ^{2}}f_{%
\frac{D-1}{2}}\left( m\Delta _{D}(u)\right) \right] ,  \label{Spm5}
\end{eqnarray}%
where the integration in the second term goes along the real axis.

The part coming from the first term in the square brackets coincides with
the corresponding function for the inertial vacuum in the Rindler
coordinates (see (\ref{SMtr3})) and for the trace of the Hadamard function
we can write%
\begin{eqnarray}
\mathrm{Tr}\left( S^{(1)}(x,x^{\prime })\right) &=&\mathrm{Tr}(S_{\mathrm{MR}%
}^{(1)}(x,x^{\prime }))+\frac{4m^{D}N}{\left( 2\pi \right) ^{\frac{D+3}{2}}}
\notag \\
&&\times \int_{-\infty }^{+\infty }du\,\frac{\left( u-\Delta \tau \right)
\sinh \left( u/2\right) }{\left( u-\Delta \tau \right) ^{2}+\pi ^{2}}f_{%
\frac{D-1}{2}}\left( m\Delta _{D}(u)\right) .  \label{S16}
\end{eqnarray}%
The coincidence limit of the second term in the right-hand side of this
formula, multiplied by $-1/2$, gives the renormalized fermionic condensate
in the form (\ref{FCalt}). A similar relation between the Feynman Green
functions of the Fulling-Rindler and Minkowski vacua for a scalar field has
been discussed in \cite{Cand76,Raje20}. In particular, in Ref. \cite{Raje20}
it was shown that the relation is a consequence of the fact that the
Minkowskian Green function is a periodic sum of the corresponding function
for the Fulling-Rindler vacuum (the thermal nature of the Minkowski vacuum
for Rindler observers). The generalization of the relation between the
Hadamard functions of scalar fields in locally Minkowski and Rindler
spacetimes, with a part of spatial dimensions compactified to a torus, is
given in \cite{Kota22}.

\end{document}